\documentclass[11pt]{article}
\usepackage[latin1]{inputenc}
\usepackage{a4wide,epsfig,amsmath,amssymb,a4wide}

\usepackage{color} 

\newcommand{\lastcorrections}%
{{
 \begin{sloppypar}
    \baselineskip -0.2in
    \tiny\bf\noindent
last corrections:\\
\end{sloppypar}
}}

\newcommand{\margincomment}[1]%
    {{%
      \marginpar{{\tiny\begin{minipage}{0.5in}
                       \begin{flushleft}
                          {#1}
                       \end{flushleft}
                       \end{minipage}
                }}
    }}

\newcommand{\myparagraph}[1]{{\smallskip\noindent{\bf #1}}}
\newcommand{\emparagraph}[1]{{\smallskip\noindent\emph{#1}}}

\newcommand{\mycase}[1]{\mbox{{\underline{Case #1}}:\/}}

\newcommand{\UU}[3]{{\bar U_{{#1},{#2},{#3}}}}
\newcommand{\PP}[4]{{\bar P_{{#1},{#2},{#3},{#4}}}}
\newcommand{\EE}[1]{{\bar E}_{#1}}

\newcommand{\calJ}{{\cal J}}

\newcommand{\braced}[1]{{ \left\{ #1 \right\} }}

\newcommand{\suchthat}{{\;:\;}}
\newcommand{\assign}{\,{\leftarrow}\,}

\newcommand{\NP}{{\mathbb{NP}}}

\newcommand{\malymax}{{\mbox{\tiny\rm max}}}

\newcommand{\pmtn}{{\mbox{\rm pmtn}}}
\newcommand{\algA}{{\sc AlgA}}
\newcommand{\algB}{{\sc AlgB}}
\newcommand{\algC}{{\sc AlgC}}

\newcommand{\ShiftBack}{{\sf ShiftBack}}
\newcommand{\Truncate}{{\sf Truncate}}
\newcommand{\load}{{\mbox{\it load}}}

\newcommand{\malyedf}{{\mbox{\tiny\rm ED}}}
\newcommand{\greedyC}{C^\malyedf}
\newcommand{\prevr}{\mbox{\it prevr}}
\newcommand{\makespan}{{C_\malymax}}

\newcommand{\RHS}{{\mbox{\rm RHS}}}
\newcommand{\Gaps}{{\mbox{\sf Gaps}}}

\newtheorem{theorem}{Theorem}

\newtheorem{lemma}{Lemma}

\newenvironment{proof}{{\noindent\bf Proof:\/}}{$\Box$\vskip 0.1in}
		{$\spadesuit$\smallskip}

\newenvironment{bigeqn*}{\large\begin{eqnarray*}}{\end{eqnarray*}}

\date{}

\begin{document}

\title{Polynomial Time Algorithms for Minimum Energy Scheduling}

\author{Philippe Baptiste%
\thanks{CNRS, LIX UMR 7161, Ecole Polytechnique 
        91128 Palaiseau, France.
         Supported by CNRS/NSF grant 17171 and ANR Alpage.} 
\and
Marek Chrobak%
\thanks{Department of Computer Science,
         University of California, Riverside, CA 92521, USA.
        Supported by NSF grants OISE-0340752, CCR-0208856 and CCF-0729071.}
\and
Christoph D\"urr%
\footnotemark[1]
}

\maketitle

\begin{abstract}
The aim of power management policies is to reduce the amount of energy
consumed by computer systems while maintaining satisfactory
level of performance.  One common method for saving energy is to
simply suspend the system during idle times. No energy is consumed
in the suspend mode.  However, the process of waking up the system
itself requires a certain fixed amount of energy, and thus suspending
the system is beneficial only if the idle time is long enough to
compensate for this additional energy expenditure.  In the specific
problem studied in the paper, we have a set of jobs with release times
and deadlines that need to be executed on a single processor.
Preemptions are allowed. The processor requires energy $L$ to be woken
up and, when it is on, it uses one unit of energy per one
unit of time.  It has been an open problem whether a schedule
minimizing the overall energy consumption can be computed in
polynomial time.  We solve this problem in positive, by providing an
$O(n^5)$-time algorithm.  In addition we provide an $O(n^{4})$-time
algorithm for computing the minimum energy schedule
when all jobs have unit length.
\end{abstract} 


\section{Introduction}


\paragraph{Power management strategies.}
The aim of power management policies is to reduce the amount of energy
consumed by computer systems while maintaining satisfactory
level of performance.  One common method for saving energy is a
\emph{power-down mechanism}, which is to simply suspend the system
during idle times.  The amount of energy used in the suspend mode
is negligible.  However, during the wake-up process the system
requires a certain fixed amount of \emph{start-up} energy, and thus
suspending the system is beneficial only if the idle time is long
enough to compensate for this additional energy expenditure. 


\myparagraph{Scheduling to minimize energy consumption.}  The
scheduling problem we study in this paper is quite fundamental. We are
given a set of jobs with release times and deadlines that need to be
executed on a single processor.  Preemptions are allowed. We assume,
without loss of generality, that, when the processor is on, it uses
one unit of energy per unit of time. The energy required to
wake up the processor is denoted by $L$. The objective
is to compute a feasible schedule that minimizes the overall energy
consumption, or to report that no feasible schedule exists.
Denoting by $E$ the energy consumption function,
this problem can be classified using Graham's notation as
$1|r_j;{\pmtn}|E$.

The question whether this problem can be solved in polynomial time was
posed by Irani and Pruhs~\cite{IraniPruhs:Algorithmic-problems}, who
write that {``\ldots Many seemingly more complicated problems in this
area can be essentially reduced to this problem, so a polynomial time
algorithm for this problem would have wide application.''}  Some
progress towards resolving this question has already been reported.
Chretienne~\cite{Chretienne:On-the-no-wait-single-machine} proved that
it is possible to decide in polynomial time whether there is a
schedule with no idle time. More recently,
Baptiste~\cite{Baptiste:min-idle-periods} showed that the problem can
be solved in time $O(n^7)$ for unit-length jobs and $L=1$.


\myparagraph{Our results.}  We solve the open problem posed by Irani
and Pruhs~\cite{IraniPruhs:Algorithmic-problems}, by providing a
polynomial-time algorithm for $1|r_j;{\pmtn}|E$.  Our algorithm is
based on dynamic programming and it runs in time $O(n^5)$. Thus not
only our algorithm solves a more general version of the problem, but
is also faster than the algorithm for unit jobs in
\cite{Baptiste:min-idle-periods}.  For the case of unit jobs (that is,
$1|r_{j};p_{j}=1|E$), we improve the running time further to $O(n^{4})$.

The paper is organized as follows. First, in Section~\ref{sec:
preliminaries}, we introduce the necessary terminology and establish
some basic properties.  Our algorithms are developed gradually in the
sections that follow.  We start with the special case of minimizing
the number of gaps for unit jobs, that is $1|r_{j};p_{j}=1;L=1|E$, for
which we describe an $O(n^4)$-time algorithm in Section~\ref{sec:
P1L1}.  Next, in Section~\ref{sec: PanyL1}, we extend this algorithm
to jobs of arbitrary length ($1|r_{j};{\pmtn};L=1|E$), increasing the
running time to $O(n^5)$.  Finally, in Section~\ref{sec: PanyLany}, we
show how to extend these algorithms to arbitrary $L$, without
affecting their running times.

We remark that although our algorithms are based on dynamic
programming, they are sensitive to the structure of the input instance
and on typical instances they are likely to run significantly faster
than their worst-case bounds.


\myparagraph{Other relevant work.}  The non-preemptive version of our
problem, that is $1|r_j|E$, can be easily shown to be $\NP$-hard in
the strong sense, even for $L=1$ (when the objective is to only
minimize the number of \emph{gaps} -- see Section~\ref{sec: preliminaries}), 
by reduction from 3-Partition~\cite[problem SS1]{GarJoh79}.

More sophisticated power management systems may involve several sleep
states with decreasing rates of energy consumption and increasing
wake-up overheads. In addition, they may also employ a method called
\emph{speed scaling} that relies on the fact that the speed (or
frequency) of processors can be changed on-line. As the energy
required to perform the job increases quickly with the speed of the
processor, speed scaling policies tend to slow down the processor
while ensuring that all jobs meet their deadlines (see
\cite{IraniPruhs:Algorithmic-problems}, for example). This problem is
a generalization of $1|r_j;{\pmtn}|E$ and its status remains open.  A
polynomial-time $2$-approximation algorithm for this problem (with two
power states) appeared in \cite{IrShGu03B}.

As jobs to be executed are often not known in advance, the on-line
version of energy minimization is of significant interest.  Online
algorithms for power-down strategies with multiple power states were
considered in \cite{IrGuSh02,IrShGu03A,AuIrSw04}.  In these works,
however, jobs are critical, that is, they must be executed as soon as
they are released, and the online algorithm only needs to determine
the appropriate power-down state when the machine is idle.  The work
of Gupta, Irani and Shukla~\cite{IrShGu03B} on power-down with speed
scaling is more relevant to ours, as it involves aspects of job
scheduling.  For the specific problem studied in our paper, $1|r_j;{\pmtn}|E$,
it is easy to show that no online algorithm can have a constant
competitive ratio (independent of $L$), even for unit jobs. We refer
the reader to \cite{IraniPruhs:Algorithmic-problems} for a detailed
survey on algorithmic problems in power management.


\section{Preliminaries}
\label{sec: preliminaries}


%
%

\myparagraph{Minimum-energy scheduling.}  
We assume that the time is discrete. More specifically, the time
is divided into unit-length intervals
$[t,t+1)$, where $t$ is an integer, called \emph{time slots} or
\emph{steps}. For brevity, we often refer to time step $[t,t+1)$ as
\emph{time step $t$}. 

An instance of the
scheduling problem $1|r_j;{\pmtn}|E$ consists of $n$ jobs, where each
job $j$ is specified by its processing time $p_{j}$, release time
$r_j$ and deadline $d_j$.  We have one processor that, at each step,
can be on or off. When it is on, it consumes energy at the rate of one
unit per time step.  When it is off, it does not consume any
energy. Changing the state from off to on (waking up) requires
additional $L$ units of energy. 

A \emph{preemptive schedule} $S$ specifies, for each
time slot, whether some job is executed at this time slot and if so,
which one.  Each job $j$ must be executed for $p_j$ time slots, and
all its time slots must be within the time interval $[r_j,d_j)$.
We say that $S$ is \emph{busy} in a given time step if it executes a job in 
this time step and that it is \emph{idle} otherwise.
A \emph{block} of a schedule $S$ is a maximal interval where $S$ is
busy. The union of all blocks of $S$
is called its \emph{support}. A \emph{gap} of $S$ is a maximal
finite interval where $S$ is idle (that is, the infinite
idle intervals before
executing the first job and after executing the last jobs are not
counted as gaps).


\begin{figure}[htbp]
\begin{center}
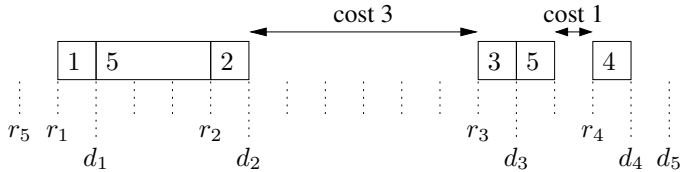
\caption{
An example of an instance of five jobs and an optimal schedule for $L = 3$.
The total energy value is $4$. Note that there are other optimal schedules
for this instance.}
\label{fig:example}
\end{center}
\end{figure}

Suppose that the input instance is feasible.  Since the energy used on
the support of all schedules that schedule all jobs is the same, it
can be subtracted from the energy function for the purpose of
minimization. The resulting function $E(S)$ is the ``wasted energy''
(when the processor is on but idle) plus $L$ times the number of
wake-ups. Formally, this can be calculated as follows.  Let
$[u_{1},t_{1}),\ldots,[u_{q},t_{q})$ be the set of all blocks of $S$,
where $u_1 < t_1 < u_2 < \ldots < t_q$.  Then
\begin{eqnarray*}
	E(S) &=& \sum_{i=2}^{q}\min\braced{u_{i}-t_{i-1},L}.
\end{eqnarray*}
(We do not charge for the first wake-up at time $u_1$, since this term
is independent of the schedule.)  Intuitively, this formula reflects
the fact that once the support of a schedule is given, the optimal
suspension and wake-up times are easy to determine: we suspend the
machine during a gap if and only if its length is at least $L$, for
otherwise it would be cheaper to keep the processor on during the gap.

Our objective is to find a schedule $S$ that meets all job deadlines
and minimizes $E(S)$. (If there is no feasible schedule, we assume
that the energy value is $+\infty$.)  Note that the special case $L=1$
corresponds to simply minimizing the number of gaps.
See Figure~\ref{fig:example} for an example.

By $C_j(S)$ (or simply $C_j$, if $S$ is understood from context) we
denote the completion time of a job $j$ in a schedule $S$.  By
$\makespan(S) = \max_j C_j(S)$ we denote the maximum completion time
of any job in $S$.  We refer to $\makespan(S)$ as the \emph{completion
time of schedule $S$}.

 
\myparagraph{Simplifying assumptions.}  Throughout the paper we assume
 that jobs are ordered according to deadlines, that is $d_{1}\leq
 \ldots\leq d_{n}$.  Without loss of generality, we also assume that
 all release times are distinct and that all deadlines are
 distinct. Indeed, if $r_{i}=r_{j}$ for some jobs $i<j$, since the
 jobs cannot start both at the same time $r_{i}$, we might as well
 increase by $1$ the release time of $j$. A similar argument applies
 to deadlines.

To simplify the presentation, we will assume that the job indexed by $1$
is a special job with minimum release time $r_1$,
$p_1 = 1$ and $d_1 = r_1+1$, that is job $1$ has
unit length and must be scheduled at its release time. (Otherwise, if
job $1$ does not satisfy these conditions, we can always add such an
extra job, released $L+1$ time slots before $r_1$. This increases each
schedule's energy consumption by exactly $L$ and does not affect the
asymptotic running time of our algorithms.)

Without loss of generality, we can also assume that the input instance
is feasible. A feasible schedule corresponds to a matching between
units of jobs and time slots, so Hall's theorem gives us the following
necessary and sufficient condition for feasibility: for all time intervals $[u,v)$,
\begin{equation}					\label{feasibility}
	\sum_{u\le r_{j}, d_{j}\le v} p_{j} \leq v-u,
\end{equation}
which in particular implies $d_j\ge r_j + p_j$ for all $j$.
It is well-known that condition (\ref{feasibility}) can be efficiently verified 
by computing the \emph{greedy earliest-deadline schedule} that at each time
step schedules the earliest-deadline pending job --
see for example \cite[p. 70]{Brucker:04:Scheduling-Algorithms} and the discussion
later in this section. 
Condition (\ref{feasibility}) will play an important role in 
correctness proofs of our algorithms.

We can also restrict our attention to schedules $S$ that satisfy the
following \emph{earliest-deadline property}: at any time $t$, either
$S$ is idle at $t$ or it schedules a pending job with the earliest
deadline. (We emphasize that this concept is more general than
the greedy earliest-deadline schedule mentioned in the 
paragraph above, because a schedule that obeys the earliest-deadline
property could be idle even if there is a pending job.)
Note that the schedule in Figure~\ref{fig:example} has this property.
In other words, once the support of $S$ is fixed, 
within the support we can schedule the jobs one by one, from left
to right, in each slot of the support executing the pending job with 
minimum deadline.
Using the standard exchange argument, any schedule can be
converted into one that satisfies the earliest-deadline property and
has the same support.
Thus, throughout the paper, we will tacitly assume (unless
explicitly noted otherwise) that all schedules we consider 
satisfy the earliest-deadline property.

We now make another observation concerning the number of gaps.
We claim that, without loss of generality, we can assume that
the optimal schedule has at most $n-1$ gaps. The argument is
quite simple: if $S$ is any optimal schedule, consider a gap $[u,v)$
and the block that follows it, say $[v,w)$. If there is no
release time in $[u,w)$, then all jobs executed in $[v,w)$ are
released before $u$, so we can shift the whole block $[v,w)$
leftwards all the way to $u$, merging two blocks. 
If $[v,w)$ was the last block,
this, clearly, decreases the cost. If $[v,w)$ is not the last
block, this change merges two gaps into one, which can only
decrease the cost. Therefore we can assume that $[u,w)$
contains a release time. As this is true for each gap in $S$, 
we conclude that the number of gaps in $S$ is at most $n-1$, as claimed.

\begin{figure}[htbp]
\begin{center}
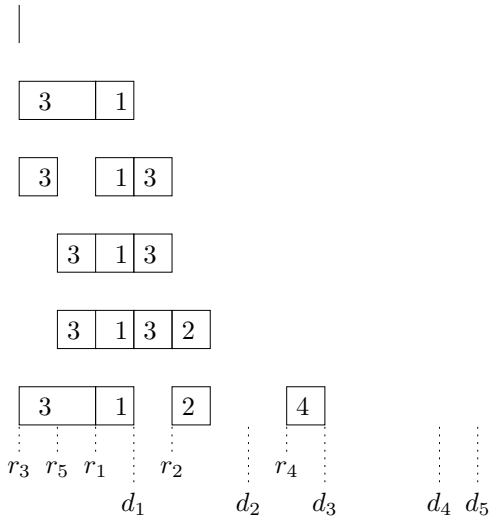
\caption{Some $(3,4)$-schedules. The first row shows the empty $(3,4)$-schedule.
In this example, $p_3 = 2$ and all other jobs have unit processing times.}
\label{fig:sk}
\end{center}
\end{figure}

\myparagraph{$(s,k)$-Schedules.}  
We will consider certain partial
schedules, that is schedules that execute only some jobs from the
instance.  For jobs $s$ and $k$, a partial schedule $S$ is called an
\emph{$(s,k)$-schedule} if it schedules all jobs $j\le k$ with $r_s
\le r_j < \makespan(S) \le d_k$. Note that different $(s,k)$-schedules
may schedule different sets of jobs. Intuitively,
as $\makespan(S)$ gets larger, then $S$
may be forced to include more jobs. See Figure~\ref{fig:sk} for illustration.

From now on, unless ambiguity arises, we will omit
the term ``partial'' and refer to partial schedules simply as
schedules.  When we say that an $(s,k)$-schedule $S$ has $g$ gaps,
in addition to the gaps between the blocks we also count the gap (if
any) between $r_s$ and the first block of $S$. 

For any $s,k$, the
empty schedule is also considered to be an $(s,k)$-schedule.
The completion time of an empty $(s,k)$-schedule is artificially
set to $r_s$. (In this convention, empty $(s,k)$-schedules,
for difference choices of $s,k$, are considered to be different
schedules.)


\paragraph{Greedy schedules.}
For any $s$, $k$, and $i$ such that $r_i\ge r_s$ and  $i\le k$, 
let $\greedyC_{s,i}$ denote 
the minimum completion time of job $i$ among all earliest-deadline 
$(s,k)$-schedules that schedule $i$. (As explained below,
$\greedyC_{s,i}$ does not depend on $k$.)
We observe that if $r_s \le r_{l} \le r_i$ then
$\greedyC_{s,i} \ge \greedyC_{l,i}$ -- simply because if we take
an earliest-deadline $(s,k)$-schedule realizing $\greedyC_{s,i}$ and remove all jobs
released before $r_{l}$, we obtain an earliest-deadline $(l,k)$-schedule that
schedules $i$. 

By $G_{s,k}$ we denote the greedy
$(s,k)$-schedule that, for each time step $t = r_s,r_s+1,...$, 
schedules the most urgent pending job. 
Note that $G_{s,k}$ may not minimize the number of gaps.
In $G_{s,k}$, the schedule of a job $i$ does not depend on any
jobs $j>i$. Therefore $C_i(G_{s,k}) = C_i(G_{s,i}) = C_i(G_{l,i})$,
for some job $l$ such that $l\le i$ and $r_s \le r_{l} \le r_i$.

\begin{figure}[ht]
\centerline{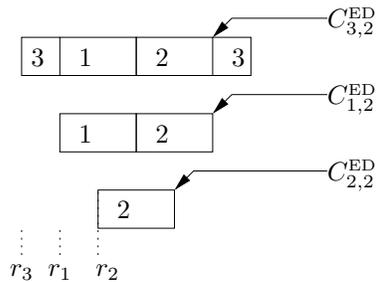}
\caption{The minimum completion time of job $i$ in an
$(s,k)$-schedule may depend on $s$. In this example,
$p_1 = p_2 = p_3 = 2$, and $k\ge 3$.}
\label{fig:bsi}
\end{figure}

The duality lemma below establishes a relation between
$\greedyC_{s,i}$ and greedy schedules. In particular, it
implies that greedy schedules are feasible (all deadlines are met).
It also shows that $\greedyC_{s,i}$ does not depend on $k$, 
justifying the omission of the subscript $k$ in the notation $\greedyC_{s,i}$.
However, $\greedyC_{s,i}$ may depend on $s$, as
illustrated in Figure~\ref{fig:bsi}.


For any times $a < b$ and a job $i$, define
\begin{eqnarray*}
	\load_i(a,b) &=& \sum_{	j\le i,\: a \le r_j < b} p_j.
\end{eqnarray*}
Thus $\load_i(a,b)$ is the total workload of the jobs 
released between $a$ and $b$ whose deadlines are at most $d_i$.


\begin{lemma}[earliest completion] \label{lem:earliest} 
For any $s$, $k$ and $i\le k$ such that $r_i\ge r_s$, we have
\begin{eqnarray}     
	\greedyC_{s,i} 
	&=& C_i(G_{s,k})    
    \;=\; \max_{\begin{subarray}{c}
					l\le i \\
					 r_s\le r_{l}\le r_i
				\end{subarray}
				} 
        \min\braced{ 
                 b \,:\, b > r_i \;\&\; b \ge r_{l} + \load_i(r_{l},b) 
            }.
			\label{eq:bi}
  \end{eqnarray}
\end{lemma}

\begin{proof}
Let {\RHS(\ref{eq:bi})} stand for the expression on the right-hand side of
(\ref{eq:bi}). It is sufficient to show that
$\greedyC_{s,i}\le C_i(G_{s,k}) \le {\RHS(\ref{eq:bi})} \le \greedyC_{s,i}$.
The inequality $\greedyC_{s,i}\le C_i(G_{s,k})$ is trivial, directly from the
definition of $\greedyC_{s,i}$. Thus it is sufficient to show the two
remaining inequalities.

We now show that $C_i(G_{s,k}) \le \RHS(\ref{eq:bi})$.
As we observed earlier, $C_i(G_{s,k})$ does not
depend on $k$ (as long as $k\ge i$, of course), by the earliest-deadline
rule, so we can assume $k=i$. Write $C_i = C_i(G_{s,i})$.
Let $l$ be the first job scheduled in $G_{s,i}$ in the block containing slot $r_i$.  
It is sufficient to show that
\begin{eqnarray}
C_i &\le& \min\braced{b \,:\, b > r_i \;\&\; b \ge r_{l} + \load_i(r_{l},b)}.
			\label{eqn: Ci < r_l+load}
\end{eqnarray}
Note that the minimum on the right-hand side of (\ref{eqn: Ci < r_l+load})
is well defined, as this set contains any $b$ that is large enough.
Thus it remains to show that for any $b$
such that  $r_i < b < C_i$ we have $b < r_{l} + \load_i(r_{l},b)$.
Indeed, consider schedule $G_{s,i}$. 
By the definition of $l$, the block containing $r_i$ starts at time $r_{l}$. 
Also, there is no idle time between $r_i$ and $C_i$. Therefore
all slots $r_{l},r_{l+1},...,b-1$ are filled with jobs $j \le i$ such that
$r_{l}\le r_j < b$. Just after scheduling
slot $b-1$, the greedy algorithm still has at least one unit of $i$ 
pending (because $i$ completes after $b$). This implies that 
$b < r_{l} + \load_i(r_{l},b)$, as claimed, completing the proof
of the inequality $C_i(G_{s,k}) \le \RHS(\ref{eq:bi})$.

Finally, we prove that ${\RHS(\ref{eq:bi})} \le \greedyC_{s,i}$.
Choose any $l \le i$ with $r_s \le r_{l}\le r_i$. Recall that
$\greedyC_{l,i} \le \greedyC_{s,i}$ (see the comments following the
definition of $\greedyC_{s,i}$). Thus, if $S$ is any earliest-deadline
$(l,i)$-schedule that schedules $i$, it is sufficient to prove that 
\begin{eqnarray}
	\min\braced{ b \,:\, b > r_i \;\&\; b \ge r_{l} + \load_i(r_{l},b) }
 &\le& C_i(S).
		\label{eqn: ec aux 1}
\end{eqnarray}
All we need to do is to show that $C_i(S)$ is a candidate for
$b$ on the left-hand side of (\ref{eqn: ec aux 1}).
That $C_i(S) > r_i$ is obvious. Further, in $S$,
at time $C_i(S)$ the least urgent job $i$ completes, so $S$ has no
pending jobs at time $C_i(S)$, which immediately implies that 
$C_i(S) \ge r_{l} + \load_i(r_{l},C_i(S))$.
\end{proof}


\paragraph{Fixed slots and segments.}
Later in the paper (in the proofs of
Lemmas~\ref{lem: unit optimal partitioning} and \ref{lem: compression}), 
we will need to show that if there exists a schedule with specific properties then there 
exists another similar schedule but with smaller completion time. For this purpose, we 
need somehow to compress the schedule, by shifting some jobs to the left, while respecting 
the release times. In order to make this formal, we now
introduce some definitions.

\begin{figure}[ht]
\centerline{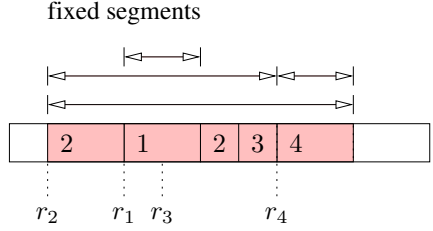}
\caption{Illustration of fixed fixed segments. Fixed slots are shaded. 
Here, $p_1 = 2$, $p_2 = 3$, $p_3=1$ and $p_4=2$.}
\label{fig: fixed slots}
\end{figure}

Let $Q$ be a schedule and let $[t',t)$ be an interval such that $Q$
is busy in all slots of $[t',t)$. We call $[t',t)$ a
\emph{fixed segment} of $Q$ if each job executed in $[t',t)$ is
released in $[t',t)$ and is completed by $Q$ in $[t',t)$.
Slots that belong to fixed segments are called \emph{fixed slots}.
By definition, if a fixed segment starts at time $u$ and $Q$ executes a job $l$ at
time $u$, then $u = r_l$. See Figure~\ref{fig: fixed slots} for
illustration.

The following lemma relates fixed segments to earliest completion times.


\begin{lemma}        \label{lem:fixed=earliest}
Consider any $s,k$, and some arbitrary $(s,k)$-schedule $S$ with
$\makespan(S) = t$. Suppose that $[u,t)$ is a fixed segment in $S$.  
Then for every job $i\le k$ that completes in this segments (that is,
$u< C_i(S) \le t$), we have $C_i(S) = \greedyC_{s,i}$.
\end{lemma}

\begin{proof}
Write $C_i = C_i(S)$.
By definition, $C_i \ge \greedyC_{s,i}$, so it is sufficient to show that
$C_i \le \greedyC_{s,i}$.

By the definition of fixed segments, $u \le r_i$.
Let $l$ be the job executed in slot $u$. Then we must have $r_{l} = u$.
Since $\greedyC_{l,i} \le \greedyC_{s,i}$, it is sufficient now to show that
$C_i \le \greedyC_{l,i}$.

The definition of fixed segments implies that all jobs executed in $[u,C_i)$
are released in $[u,C_i)$. Since, by our convention, $S$ has the earliest-deadline
property, $S$ and $G_{l,k}$ are actually identical in $[u,C_i)$, so
$C_i = C_i(G_{l,k}) = \greedyC_{l,k} = \greedyC_{l,i}$,
where the second equation follows from Lemma~\ref{lem:earliest} and the
last one from $i\le k$.
\end{proof}


\begin{figure}[ht]
\centerline{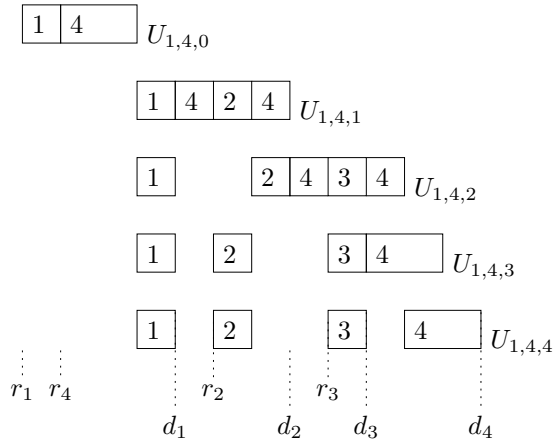}
\caption{The value $U_{s,k,g}$ is non-decreasing in $g$.
Here, $p_1 = p_2 = p_3 = 1$ and $p_4 = 2$.}
\label{fig:Uskg}
\end{figure}

\myparagraph{An outline of the algorithms.} 
For any $s= 1,...,n$, $k = 0,...,n$, and $g = 0,...,n-1$, define 
$U_{s,k,g}$ as the maximum
completion time of an $(s,k)$-schedule with at most $g$ gaps
(see Figure~\ref{fig:Uskg} for illustration). By an argument similar to the one 
given earlier in this section, we only need to consider values $g\le n-1$,
because $U_{s,k,g} = U_{s,k,n-1}$ for $g \ge n$.
  

Our algorithms consist of two stages. The first stage is to
compute the table $U_{s,k,g}$, using dynamic programming.
Note that from this table we can
determine the minimum number of gaps in the (complete) schedule:
the minimum number of gaps is
equal to the smallest $g$ for which $U_{1,n,g}> \max_jr_j$.
The algorithm computing $U_{s,k,g}$ for unit jobs is
called {\algA} and the one for arbitrary length jobs is called {\algB}.

In the second stage, described in 
Section~\ref{sec: PanyLany} and called {\algC}, we use the table $U_{s,k,g}$ to 
compute the minimum energy schedule. In other words, we show that
the problem of computing the
minimum energy reduces to computing the minimum number of gaps.  This
reduction, itself, involves again dynamic programming.

When presenting our algorithms, we will only show how to compute the
minimum energy value. The algorithms can be modified in a
straightforward way to compute the actual optimum schedule, without
increasing the running time.  (In fact, we explain how to construct
such schedules in the correctness proofs.)


%

\section{Main Idea}
\label{sec: main idea}

This section is quite informal, and its purpose is to explain the
thought process leading to the design of our algorithms. The basic
principle is what we will refer to as the \emph{inversion method} 
for speeding dynamic programming algorithms.


\paragraph{The inversion trick.}
The idea is this. Imagine you have a dynamic programming algorithm that
tabulates a function
\begin{eqnarray*}
\lambda(x) &=& \max\braced{ y \suchthat \Pi(x,y) },
\end{eqnarray*}
where $\Pi(x,y)$ is some predicate and $\lambda(x)$ is non-increasing with $x$.
As usual in dynamic programming, only one value of $\lambda(x)$, say $\lambda(x_0)$,
is actually needed to compute the
desired solution, but all values need to be tabulated. 
($\Pi(x,y)$ would typically depend recursively on some
$\lambda(z)$, for some $z's$ ``smaller" than $x$.) Suppose
that the range of $x$ is large, while the number of possible values 
$y$ is small. Then instead of computing $\lambda(x)$ we can tabulate its 
``inverse"
\begin{eqnarray*}
\delta(y) &=& \max\braced{ x \suchthat \Pi(x,y) },
\end{eqnarray*}
and then compute $\lambda(x_0)$ from $\delta(y)$'s using binary search. Since 
there are fewer $y$'s than $x$'s, this is likely to lead to a faster algorithm.


\paragraph{The case of unit-length jobs.}
What does it have to do with our algorithms? The starting point here is
the algorithm by Baptiste~\cite{Baptiste:min-idle-periods} 
for minimizing the number of gaps
for unit-length jobs (in our notation, $1|r_j;p_j = 1;L=1|E$).
This algorithm tabulates the following function:
\begin{description}
\item{$\Gaps(k,u,v) =$} the minimum number of gaps for the jobs numbered
		$1,...,k$ whose release times are in the interval $[u,v)$.
\end{description}
(To be more precise, in \cite{Baptiste:min-idle-periods}, non-empty
idle periods starting at $u$ or ending at $v$, if any, are also
counted as gaps.) Baptiste~\cite{Baptiste:min-idle-periods} 
achieved running time $O(n^7)$ by showing that $u$ and $v$ can be
chosen from $O(n^2)$-size ranges, and by giving a recurrence for
$\Gaps(k,u,v)$ that can be evaluated in time $O(n^2)$.

Some speed-up of Baptiste's algorithm can be achieved by
observing that one can assume, without loss of generality, that
$u = r_s$, for some job $s$, and that $v = r_i + q$, for some job $i$ and
integer $-n \le q \le n$. A similar idea improves the
time to evaluate $\Gaps(k,u,v)$ to $O(n)$. This results in an
$O(n^5)$-time algorithm.

To improve the time further to $O(n^4)$, we apply the inversion method.
Since the $v$'s range over a set of size $O(n^2)$ and 
$\Gaps(k,u,v)$ takes only $O(n)$ values, we can tabulate 
the function $U(u,k,g)$ defined as the maximum $v$ 
for which there is a schedule with completion time $v$
that schedules jobs numbered $1,...,k$ whose release times
are in $[u,v)$. As $u=r_s$, for some $s$, this is
exactly our table $U_{s,k,g}$. 
This reduces the table size to $O(n^3)$.
We emphasize that this does not automatically give an
improvement to $O(n^4)$, since one still needs to design an
appropriate recurrence for $U_{s,k,g}$ that can be 
evaluated in time $O(n)$, which is quite non-trivial. 
We give such a recurrence in Section~\ref{sec: P1L1}.


\paragraph{Arbitrary length jobs.}
Ignoring the issue of the running time, one can apply Baptiste's algorithm
to arbitrary jobs by simply dividing each job $j$ into $p_j$ unit-length
jobs with release times $r_j$ and deadlines $d_j$.
We can then rewrite the dynamic programming function as:
\begin{description}
\item{$\Gaps(k,p,u,v) =$} the minimum number of gaps for the jobs numbered
		$1,...,k$ whose release times are in the interval $[u,v)$, with the 
		length of job $k$ changed to $p_k \assign p$.
\end{description}
This table's size is not polynomial in $n$ anymore, because of $p$.
Since $\Gaps(k,p,u,v)$ takes only $O(n)$ values, we can apply the inversion
trick again, but this time computing the value of $p$.
The resulting function is $P(u,k,g,v)$, equal (roughly) to the minimum
amount of job $k$ required to achieve $g$ gaps in the interval
$[u,v)$. As before, we can assume that $u=r_s$, for some $s$.
We show later in the paper that we can also assume that $v = r_l$, for some $l$.
This gives rise to the table $P_{s,k,g,l}$ introduced
in Section~\ref{sec: PanyL1}. This table has size only $O(n^4)$.
The recurrence for
$P_{s,k,g,l}$ is, unfortunately, quite complicated and it involves
also table $U_{s,k,g}$ -- see Section~\ref{sec: PanyL1} for a
complete description.


\section{Minimizing the Number of Gaps for Unit Jobs}
\label{sec: P1L1}

%
%
%
%

In this section we give an $O(n^4)$-time algorithm for minimizing the
number of gaps for unit jobs, that is for $1|r_j;p_j = 1;L=1|E$. 
Recall that we assume all release times to be different and
all deadlines to be different. With this assumption, it is easy to see
that there is always a feasible schedule, 
by scheduling every job at its release time.

As described in the previous section, the general idea of the algorithm
is to compute all values of the function $U_{s,k,g}$ using dynamic
programming. Before stating the algorithm, we establish some
properties of $(s,k)$-schedules.


\paragraph{Some properties of $(s,k)$-schedules.}
For some $(s,k)$-schedules, their completion time can be 
increased, while preserving the number of gaps,
simply by appending an additional job or by moving job $k$ to the end. 
Such schedules are ``wasteful", in the sense that
they cannot possibly realize $U_{s,k,g}$. This motivates the following 
definition.

An $(s,k)$-schedule $S$ is called \emph{frugal}
if it satisfies the following properties:
\begin{description}
	
\item{(f1)} There is no job $j\le k$ with $r_j = \makespan(S)$, and

\item{(f2)} Suppose that $S$ schedules job $k$ and $\makespan(S) < d_k$.
Then either 
(i) $k$ is scheduled last in $S$
(at time $\makespan(S)-1$) and the last block contains at least one
job other than $k$, or 
(ii) $k$ is scheduled inside a block (that is, $k$ is not the first nor the
last job in a block). 

\end{description}

\begin{figure}[ht]
\centerline{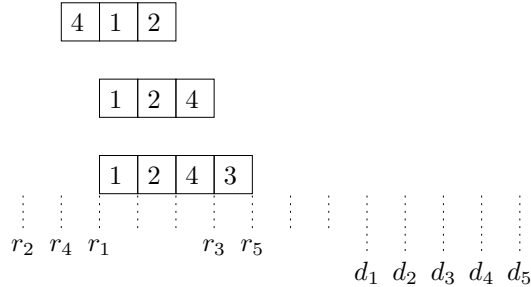}
\caption{Three $(2,4)$-schedules. The first one does not satisfy (f2).
The second schedule does not satisfy (f1) because it ends at $r_3$.
The last one is a frugal $(2,4)$-schedule, but not a frugal $(2,5)$-schedule.}
\label{fig:frugal}
\end{figure}

See Figure~\ref{fig:frugal} for illustration.


Obviously, if $\makespan(S) = d_k$, then, by the assumption about
different deadlines, job $k$ must be scheduled last in $S$.
But in this case, even if $S$ is frugal,
the last block may or may not contain jobs other than $k$.


\begin{lemma}[frugality]\label{lem: unit frugal schedules}
Fix some $s,k,g$, and let $S$ be an
$(s,k)$-schedule that realizes $U_{s,k,g}$, that is 
$S$ has at most $g$ gaps and $\makespan(S) = U_{s,k,g}$.
Then $S$ is frugal.
\end{lemma}

\begin{proof}
The proof is quite simple. If $S$ violates (f1) then 
we can extend $S$ by scheduling
$j$ at $\makespan(S)$, obtaining a new $(s,k)$-schedule
with at most $g$ gaps and larger completion time,
which contradicts the optimality of $S$.

Next, assume that $S$ satisfies condition (f1), but not (f2). 
We have two cases. Suppose first that $k$ is the
last job in $S$. Then it is not possible that
$k$ is the only job in the last block of $S$, for then we
could move $k$ to $d_k-1$, without increasing the number of
gaps but increasing the completion time. 
The other case is that $k$ is not last in $S$. If
$k$ were either the first or last job in its block,
we could reschedule $k$ at time $\makespan(S)$, 
without increasing the number of gaps and increasing
the completion time. (By condition (f1), this
is a correct $(s,k)$-schedule.) Thus in both cases we
get a contradiction with the optimality of $S$.
\end{proof}

We now make some observations that follow from the lemma above.
First, we claim that, for any fixed $s$ and $g$,
the function $k \to U_{s,k,g}$ is
non-decreasing. Indeed, suppose that $S$ is an $(s,k)$-schedule
that realizes $U_{s,k,g}$. By the lemma above, we can assume
that $S$ is frugal. If $r_{k+1}\ge \makespan(S) = u$,
then $S$ is itself a valid $(s,k+1)$-schedule.
If $r_{k+1} < u$, then we can extend $S$ by
scheduling job $k+1$ at time $u$, obtaining
a new schedule $S'$. By the frugality of $S$, no job
$j\le k$ is released at time $u$. Also,
$u \le d_k < d_{k+1}$, so $S'$ is a valid $(s,k+1)$-schedule,
it has the same number of gaps as $S$, and
$\makespan(S') > \makespan(S)$.

Further, we also claim that, for any fixed $k$ and $s$,
the function $g \to U_{s,k,g}$
is strictly increasing as long as $U_{s,k,g} < d_k$.
For suppose that $S$ is a (frugal) schedule that realizes
$U_{s,k,g}< d_k$. If there is a job $j\le k$ with
$U_{s,k,g} \le r_j < d_k$, then in fact, by frugality,
$U_{s,k,g} < r_j$. Choose such a $j$ with
minimum $r_j$ and extend $S$ by scheduling $j$ at
$r_j$. The new schedule $S'$ is an $(s,k)$-schedule, it
has one more gap than $S$, and $\makespan(S') > \makespan(S)$.
Else, suppose that such $j$ does
not exist. In particular, $r_k < U_{s,k,g}$, so
$S$ schedules $k$. Let $S'$ be the schedule obtained
from $S$ by moving $k$ to time $d_k-1$, so that
$\makespan(S') = d_k > \makespan(S)$.
$S'$ is an $(s,k)$-schedule. By the
frugality condition (f2) of $S$, either $k$
is the last job in the last block, or it is an
internal job of another block. In both cases
$S'$ has only one more gap than $S$.

\begin{figure}[ht]
\centerline{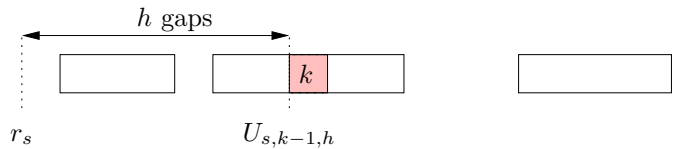}
\caption{The idea of Lemma~\ref{lem: unit optimal partitioning}.}
\label{fig:partitioning}
\end{figure}

We now show a decomposition property that leads to a dynamic program -- 
see Figure~\ref{fig:partitioning} for illustration. 
The basic idea is that, by the earliest-deadline property, the time slot
$t$ where job $k$ (the least urgent job) is executed divides the schedule 
into schedules of two disjoint sub-instances, one including jobs released 
before $t$ and the other including jobs released after. 


\begin{lemma}[partitioning]\label{lem: unit optimal partitioning}
Let $S$ be an $(s,k)$-schedule that realizes $U_{s,k,g}$
and schedules job $k$, but not as the last job.
Let $t$ be the time at which $S$ schedules job $k$, 
and let $h$ be the number of gaps in $S$ in the
interval $[r_s,t)$. Then $t = U_{s,k-1,h}$.
\end{lemma}

\begin{proof}
By Lemma~\ref{lem: unit frugal schedules}, $S$ is frugal.
Denote $v = U_{s,k-1,h}$. Clearly, by the earliest-deadline
property, no jobs $j < k$ released in $[r_s,t)$ are 
pending at time $t$. So the segment of $S$ in $[r_s,t)$ is
an $(s,k-1)$-schedule with $h$ gaps, implying that $v\ge t$.
Thus it suffices now to show that $v \le t$.  Towards
contradiction, suppose that $v>t$ and let $R$ be an $(s,k-1)$-schedule
that realizes $U_{s,k-1,h}$, that is $R$ has at most $h$
gaps and $\makespan(R) = v$.  We consider two cases.

\noindent
\mycase{1} $R$ schedules all jobs $j<k$ with $r_s\le r_j< t$ in the
interval $[r_{s},t+1)$.  We can then modify $S$ as follows: Reschedule $k$
at time $u = \makespan(S)$ and replace the segment $[r_s,t+1)$ of $S$
by the same segment of $R$.  Let $S'$ be the resulting schedule.  The
earliest deadline property of $S$ implies that there is no job $j<k$
released at time $t$. By this observation and the case condition, $S'$
is an $(s,k)$-schedule.  Also, no matter whether $R$ is idle at $t$ or
not, $S'$ has at most $h$ gaps in the segment $[r_{s},t+1)$, and
therefore at most $g$ gaps in total.  We thus obtain a contradiction
with the choice of $S$, because $\makespan(S') = u+1 > \makespan(S)$.

\noindent
\mycase{2} $R$ schedules some job $j<k$ with $r_s\le r_j< t$ strictly
after $t$.  In this case, we claim that there is an $(s,k-1)$-schedule
$R'$ (not necessarily frugal)
with at most $h$ gaps and $\makespan(R') = t+1$.  We could then
again obtain a contradiction by proceeding as in Case~1.

In Section~\ref{sec: preliminaries} we defined the concept
of fixed segments in a schedule. For unit jobs, the definition of fixed
segments becomes very simple: they consist of jobs scheduled at their
release times. This follows from
the assumption that all release times are different.
In particular, a slot $z$ of $R$ is \emph{fixed} if
the job scheduled at time $z$ is released at $z$.

Let $[w,v)$ be the last block of $R$.
To obtain $R'$, we gradually ``compress'' $R$, according to the
procedure below (see Figure~\ref{fig:compression}).

If the slot $v-1$ is fixed, then we simply remove it, replacing $R$ by
its segment in $[r_{s},v-1)$.  The result is still an
$(s,k-1)$-schedule, even though it is not frugal.  This schedule has
completion time strictly smaller than $v$, but not less than $t+2$
because, by the case assumption, strictly after time $t$ it schedules a
job $j<k$ with $r_s\le r_j< t$, and $j$'s execution slot is not fixed.


\begin{figure}[ht]
\centerline{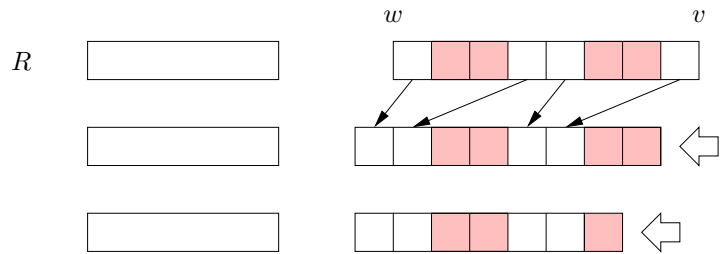}
\caption{Illustration of the compression. Fixed slots are shown shaded.
The first step corresponds to the sub-case of Case~2 when slot $v-1$ is not fixed and
some jobs are shifted. The second case corresponds to the sub-case of Case~2 when 
slot $v-1$ is fixed and the last job is removed from the schedule.}
\label{fig:compression}
\end{figure}


The other case is when the slot $v-1$ is not fixed.  Now for each
non-fixed slot in $[w,v)$, move the job in this slot to the previous
non-fixed slot. The job from the first non-fixed slot will move to
slot $w-1$.  By the assumption about distinct release times, this
operation will not move a job before its release time. It also
preserves fixed slots, while some non-fixed slots, including the empty
slot $w-1$, might become fixed.  The last block now ends one unit
earlier, and either it starts one unit earlier or is merged with the
second last block. After this operation, $R$ remains an
$(s,k-1)$-schedule with at most $h$ gaps. If $\makespan(R) = t+1$, we
let $R' = R$, otherwise we continue the process.
\end{proof}


\paragraph{Outline of the algorithm.}
As explained in the previous section, the algorithm computes the table
$U_{s,k,g}$.  The crucial idea here is this: Let $S$ be an
$(s,k)$-schedule that realizes $U_{s,k,g}$, that is $S$ has at most $g$ gaps
and $\makespan(S)$ is maximized. If $S$ does not schedule $k$, then
$S$ is an $(s,k-1)$-schedule, so $U_{s,k,g} = U_{s,k-1,g}$.  If $S$
schedules $k$ as the last job, then either $U_{s,k,g} = \makespan(S) =
d_k$ or the last block contains jobs other than $k$, in which case
the part of $S$ before $k$ is an $(s,k-1)$-schedule with the same number of 
gaps $g$, implying that $U_{s,k,g} = U_{s,k-1,g}+1$.  The most interesting
case is when $S$ schedules $k$ not as the last job, say at time
$t$. By frugality, $k$ is neither the first nor the last job in its
block.  Denote $u = U_{s,k,g}$. We show that, without loss of
generality, there is a job $l$ released and scheduled at time
$t+1$. Further, the segment of $S$ in $[r_s,t)$ is an
$(s,k-1)$-schedule with completion time $t$, the segment of $S$ in
$[t+1,u)$ is an $(l,k-1)$-schedule with completion time $U_{s,k,g}$,
and the total number of gaps in these two schedules is at most
$g$. Denoting by $h$ the number of gaps of $S$ in the interval
$[r_s,t)$, we conclude that $U_{s,k,g} = U_{l,k-1,g-h}$, and by
Lemma~\ref{lem: unit optimal partitioning} we also have $t =
U_{s,k-1,h}$, leading naturally to a recurrence relation for this
case.


\paragraph{Algorithm~{\algA}.}
The algorithm computes all values $U_{s,k,g}$, for
$s=1,...,n$,  $k = 0,...,n$, and $g = 0,...,n-1$, using dynamic programming.  
The minimum
number of gaps for the input instance is equal to the smallest $g$ for
which $U_{1,n,g} > \max_j r_j$.

The values $U_{s,k,g}$ will be stored in the table $\UU{s}{k}{g}$. To
explain how to compute this table, we give the appropriate recurrence
relation.  

For the base case $k = 0$, we let $\UU{s}{0}{g} \assign r_s$ for all
$s$ and $g$. For $k\ge 1$, we proceed like this.  If $r_k < r_s$ then
$\UU{s}{k}{g} = \UU{s}{k-1}{g}$.  Otherwise we have $r_k \ge r_s$, in
which case $\UU{s}{k}{g}$ is defined recursively as follows:
\begin{eqnarray}
\UU{s}{k}{g} &\assign&
   \max
	\left\{\begin{array}{lcl}
       \phantom{\max \{} \UU{s}{k-1}{g}  
                && \textrm{if } \UU{s}{k-1}{g} < r_k
\\
       \phantom{\max \{} \UU{s}{k-1}{g} + 1 
                && \textrm{if } \UU{s}{k-1}{g} \geq r_k
\\
\multicolumn3l {
\max \{
       \UU{l}{k-1}{g-h} \::\: l<k, h\leq g, 
                r_k < r_l = \UU{s}{k-1}{h}+1 \}
}
\\
       \phantom{\max \{} d_k 
                && \textrm{if $g\ge 1
							\ \&\  (\, r_j < \UU{s}{k-1}{g-1} \;\forall j<k \,)$}
           \end{array}
          \right.
                \label{eqn:A}
\end{eqnarray}
Note that in the third option the variables $l$ and $h$ are dependent:
if we fix the value of one, then the other one's value is fixed as
well (or it does not exist). In the maximum for this option, we assume
that its value is $-\infty$, if there are no $l,h$ that satisfy its
condition. Note also that the maximum (\ref{eqn:A}) is well-defined,
because either the first or the second option applies.

In the remainder of this section we justify the correctness
of the algorithm and analyze its running time. The first 
lemma establishes the feasibility and optimality of
the values $\UU{s}{k}{g}$ computed by Algorithm~{\algA}.
The main idea was explained earlier in this section and is
quite simple, but the formal proof is rather involved. This is
partially due to the fact that we carry out the
feasibility and optimality proofs jointly, because in
some situations the feasibility of some $(s,k)$-schedules we
construct depends on frugality (and thus also, indirectly, on
optimality) of its $(s',k-1)$-sub-schedules.

\begin{figure}[htb]
  \centerline{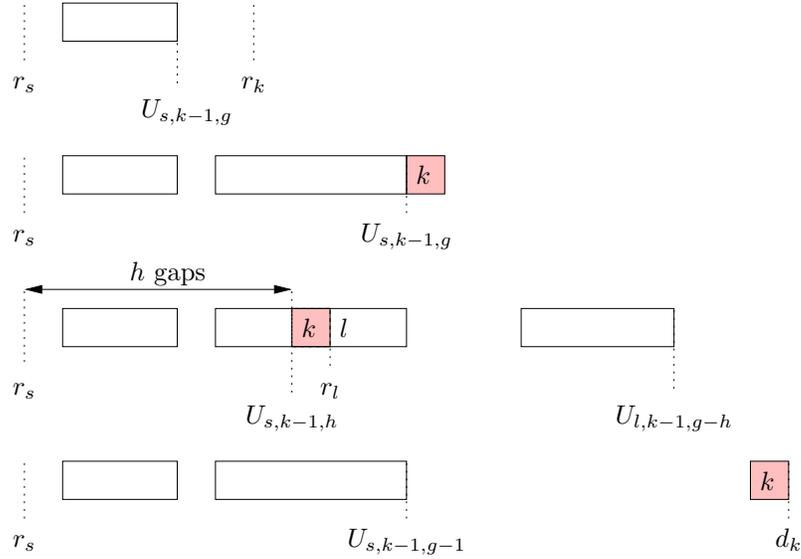}
  \caption{Illustration of the cases in
	in the proof of Lemma~\ref{lem: correctness A}.}
  \label{fig: correctness A}
\end{figure}


\begin{lemma}[correctness of {\algA}]\label{lem: correctness A}
Algorithm~{\algA} correctly computes the values $U_{s,k,g}$,
that is $\UU{s}{k}{g} = U_{s,k,g}$  for all $s = 1,...,n$, $k = 0,...,n$,  
and $g = 0,...,n-1$.
\end{lemma}

\begin{proof}
It is sufficient to show that the following two claims hold:
\begin{description}
	\item{\emph{Feasibility:} }
	For any choice of indices $s,k,g$, there is
	an $(s,k)$-schedule $S_{s,k,g}$ with $\makespan(S_{s,k,g}) = \UU{s}{k}{g}$ 
	and at most $g$ gaps.
	\item{\emph{Optimality:} }
	For any choice of indices $s,k,g$,	if $Q$ is any $(s,k)$-schedule
	with at most $g$ gaps then $\makespan(Q) \le \UU{s}{k}{g}$.
\end{description}

The proof is by induction on $k$.
Consider the base case first, for $k=0$.
To show feasibility, we take $S_{s,0,g}$ to
be the empty $(s,0)$-schedule, which is trivially feasible and (by
our convention) has completion time $r_s = \UU{s}{0}{g}$. 
The optimality condition follows from the fact that any $(s,0)$-schedule 
is empty and thus has completion time $r_s$.

Suppose now that the feasibility and optimality conditions hold
for $k-1$. We will show that they hold for $k$ as well.

\medskip 
\emph{Feasibility proof.}  By the inductive assumption, for
any $s'$ and $g'$ we have a schedule $S_{s',k-1,g'}$ with completion
time $\UU{s'}{k-1}{g'} = U_{s',k-1,g'}$.  By Lemma~\ref{lem: unit
  frugal schedules}, $S_{s',k-1,g'}$ is frugal.  The construction of
$S_{s,k,g}$ depends on which expression realizes the value of
$\UU{s}{k}{g}$. If $r_k<r_s$, then any $(s,k-1)$-schedule is also a $(s,k)$-schedule 
and therefore, by the inductive assumption, $S_{s,k,g} = S_{s,k-1,g}$
is an $(s,k)$-schedule with completion time $\UU{s}{k}{g}$.  
From now on assume $r_k\geq r_s$.

\noindent
\mycase{1} If $\UU{s}{k}{g}=\UU{s}{k-1}{g}$ and $\UU{s}{k-1}{g} < r_k$, then 
we simply take $S_{s,k,g} = S_{s,k-1,g}$. Therefore,
from the inductive assumption, and the inequality we get that $S_{s,k,g}$ is an
$(s,k)$-schedule with completion time $\UU{s}{k}{g}$.

\noindent
\mycase{2}
If $\UU{s}{k}{g} = \UU{s}{k-1}{g}+1$ and $\UU{s}{k-1}{g} \geq r_k$,
then let $S_{s,k,g}$ be the schedule obtained from $S_{s,k-1,g}$ by
appending to it job $k$ scheduled at time $u = \UU{s}{k-1}{g}$. By the
frugality of $S_{s,k-1,g}$, there is no job $j\le k$ with $r_j =
u$. We also have $u < d_k$, which follows from $u\le d_{k-1}$ and the
assumption about distinct deadlines.  Therefore $S_{s,k,g}$ is an
$(s,k)$-schedule with completion time $u+1 = \UU{s}{k}{g}$.

\noindent
\mycase{3}
Next, suppose that $\UU{s}{k}{g} = \UU{l}{k-1}{g-h}$, for some
$1\le l < k$, $0 \le h \le g$, that satisfy $r_k < r_l =
\UU{s}{k-1}{h}+1$.  The schedule $S_{s,k,g}$ is obtained by scheduling
all jobs $j<k$ released between $r_s$ and $r_l-1$ using $S_{s,k-1,h}$,
scheduling all jobs $j<k$ released between $r_l$ and
$\UU{l}{k-1}{g-h}-1$ using $S_{l,k-1,g-h}$, and scheduling job $k$ at
$r_l-1$. By the frugality of $S_{s,k-1,h}$, there is no job $j<k$ with
$r_j = r_l-1$.  Thus $S_{s,k,g}$ is an $(s,k)$-schedule with completion
time $\UU{s}{k}{g}$ and at most $g$ gaps.

\noindent
\mycase{4}
Finally, suppose that $\UU{s}{k}{g} = d_k$, $g \ge 1$, and $\max_{j<k}
r_j < \UU{s}{k-1}{g-1}$. Let $S_{s,k,g}$ be the schedule obtained from
$S_{s,k-1,g-1}$ by adding to it job $k$ scheduled at $d_k-1$.  The
case condition implies that no jobs $j < k$ are released between
$\UU{s}{k-1}{g-1}$ and $d_k-1$. By the assumption about different
deadlines, we also have $\UU{s}{k-1}{g-1} < d_k$.  Therefore
$S_{s,k,g}$ is an $(s,k)$-schedule with completion time $d_k =
\UU{s}{k}{g}$ and it has at most $g$ gaps, since adding $k$ can add at
most one gap to $S_{s,k-1,g-1}$.

\medskip 
\emph{Optimality proof.}   Let $Q$ be an
$(s,k)$-schedule with at most $g$ gaps and completion time
$u = \makespan(Q)$. We can assume that $Q$ realizes
$U_{s,k,g}$, that is, $u = U_{s,k,g}$.  Without loss of
generality, we can also assume that $Q$ has the earliest-deadline property
and is frugal. In particular, this implies that no job $j\le k$ is
released at time $u$.  We prove that $u \le \UU{s}{k}{g}$ by analyzing
several cases.


\noindent
\mycase{1} $Q$ does not schedule job $k$. In this case $Q$ is an
$(s,k-1)$-schedule with completion time $u$, so, by induction, we have
$u \le \UU{s}{k-1}{g} \le \UU{s}{k}{g}$.

In all the remaining cases, we assume that $Q$ schedules $k$.
Obviously, this implies that $r_s \le r_k < u$.


\noindent
\mycase{2} $Q$ schedules $k$ as the last job and $k$ is not the only
job in its block.  Let $u' = u-1$, and define $Q'$ to be $Q$
restricted to the interval $[r_{s},u')$. Then $Q'$ is an
$(s,k-1)$-schedule with completion time $u'$ and at most $g$ gaps, so
$u'\le \UU{s}{k-1}{g}$, by induction.  
Since $k$ is executed at time $u'$ in $Q$, we
have $r_k\le u' \le \UU{s}{k-1}{g}$, so the second option of 
the maximum (\ref{eqn:A}) is applicable.  
Therefore $u = u' + 1 \le \UU{s}{k-1}{g} + 1 \le \UU{s}{k}{g}$.


\noindent
\mycase{3} $Q$ schedules $k$ and $k$ is not the last job.  Suppose
that $k$ is scheduled at time $t$.  By the frugality of $Q$, $k$ is
neither the first nor last job in its block.  Since $Q$ satisfies the
earliest-deadline property, no job $j < k$ is pending at time $t$, and
thus $Q$ schedules at time $t+1$ the job $l<k$ with release time
$r_{l}=t+1$ (see the third case in Figure~\ref{fig: correctness A}).

By Lemma~\ref{lem: unit optimal partitioning} and induction, $t =
U_{s,k-1,h} = \UU{s}{k-1}{h}$ for some $h\le g$. Then the conditions
of the third option in (\ref{eqn:A}) are met: $l < k$, $h\le g$, and
$r_k < r_l = \UU{s}{k-1}{h}+1$. Let $Q'$ be the segment of $Q$ in
$[r_l,u)$. Then $Q'$ is an $(l,k-1)$-schedule with completion time $u$
and at most $g-h$ gaps, so by induction we get $u \le \UU{l}{k-1}{g-h}
\le \UU{s}{k}{g}$, completing the argument for Case~3.


\noindent
\mycase{4} $Q$ schedules $k$ as the last job and $k$ is the only job
in its block.  If $u = r_s+1$ then $k = s$ and the second option
of (\ref{eqn:A}) is applicable (because $r_s\le \UU{s}{s-1}{g}$), 
so we have $u = r_s+1 \le
\UU{s}{s-1}{g}+1 \le \UU{s}{s}{g}$. Thus we can assume now that
$u > r_s+1$, which, together with the case condition, implies that $g > 0$.
By the case assumption and the frugality of $Q$, we can also
assume that $u = d_k$. (To see why, observe that in the definition of
frugality, in part (f2), neither (i) nor (ii) applies to $Q$.)

Let $u'$ be the earliest time $u'\geq r_s$ such that $Q$ is idle in
$[u',d_{k}-1)$. Then, by the feasibility of $Q$, 
$\max_{j < k} r_j < u'$ and the segment of $Q$ in
$[r_s,u')$ is an $(s,k-1)$-schedule with at most $g-1$ gaps.  So, by
induction, we get $u' \le \UU{s}{k-1}{g-1}$.  Thus the last option in
(\ref{eqn:A}) applies and we get $u = d_k = \UU{s}{k}{g}$.
\end{proof}


\begin{theorem}\label{thm: algorithm A}
Algorithm~{\algA} correctly computes the optimum solution for
$1|r_j;p_j = 1; L=1|E$, and it can be implemented in time $O(n^4)$.
\end{theorem}

\begin{proof}
  The correctness of Algorithm~{\algA} follows from Lemma~\ref{lem:
    correctness A}, so it is sufficient to give the running time
  analysis.  There are $O(n^3)$ values $\UU{s}{k}{g}$ to be computed.
  For fixed $s,k,g$, the first two choices in the maximum
  (\ref{eqn:A}) can be computed in time $O(1)$ and the last choice in
  time $O(n)$.  In the third choice we maximize only over pairs $(l,h)$
  that satisfy the condition $r_l = \UU{s}{k-1}{h}+1$, and thus we
  only have $O(n)$ such pairs. Further, since the values of
  $\UU{s}{k-1}{h}$ increase with $h$, we can determine all these pairs
  in time $O(n)$ by searching for common elements in two sorted lists:
  the list of release times, and the list of times $\UU{s}{k-1}{h}+1$,
  for $h=0,1,...,n$.  Thus each value $\UU{s}{k}{g}$ can be computed
  in time $O(n)$, and we conclude that the overall running time of
  Algorithm~{\algA} is $O(n^4)$.
\end{proof}



\section{Minimizing the Number of Gaps for Arbitrary Jobs}
\label{sec: PanyL1}

In this section we give an $O(n^5)$-time algorithm for minimizing the
number of gaps for instances with jobs of arbitrary lengths, that is
for the scheduling problem $1|r_j;{\pmtn}; L=1|E$.  

\begin{figure}[ht]
\centerline{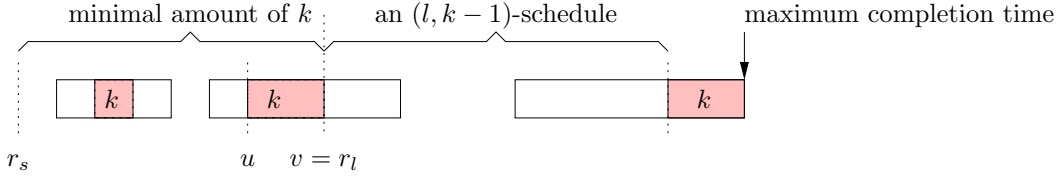}
\caption{The fundamental idea of Algorithm~\algB.}
\label{fig:structure_of_S}
\end{figure}

As in Algorithm~{\algA}, we focus on computing the function $U_{s,k,g}$.
The new recurrence relations for $U_{s,k,g}$ are significantly more
involved than in Algorithm~{\algA}, but the fundamental principle is quite
intuitive (see Figure~\ref{fig:structure_of_S}): 
Imagine an $(s,k)$-schedule $S$ with at most $g$ gaps 
that maximizes completion time. If the last internal execution
interval of $k$ in $S$ ends at $v$, then, 
by the earliest-deadline property we have $v=r_l$, for some job $l<k$.
Further, the segment of $S$ in $[r_s,v)$
must have a minimum number of units of $k$, for otherwise these
units could be moved to the end of $S$ increasing its completion time.
We represent this minimum number of units of $k$ in $[r_s,v)$ by
another function $P_{s,k,h,l}$, where $h$ is the number of gaps of $S$
in $[r_s,v)$. On the other hand, the segment of $S$ starting at $v$
consists of an $(l,k-1)$-schedule followed by some number of units of
$k$. This structure of $S$ allows us to express $U_{s,k,g}$ in terms 
of $P_{s,k,h,l}$ and $U_{l,k-1,g-h}$.

The above intuition, although fundamentally correct, glosses over
some important technical issues and ignores some special cases (for
example, when $S$ completes at $d_k$). To formalize this idea we
need to establish some properties of optimal schedules.
We proved some results about the structure
of optimal schedules for unit jobs in the previous section; we now extend 
those results to jobs of arbitrary length.


\paragraph{Frugal $(s,k)$-schedules.}
Given a schedule $S$, by an \emph{execution interval} $[u,v)$ of job $k$ we
mean an inclusion-wise maximal time interval where $S$ executes $k$
(that is, $k$ is scheduled in each time unit inside $[u,v)$ but is not
scheduled at times $u-1$ and $v$).

An $(s,k)$-schedule $S$ is called \emph{frugal} if it satisfies the
following properties:
\begin{description}
	
\item{(f1)} There is no job $j\le k$ with $r_j = \makespan(S)$, and

\item{(f2)} Suppose that $\makespan(S) < d_k$ and $S$ schedules job $k$. 
Let $[u,v)$ be an execution interval of job $k$.
Then the slot $u-1$ is not idle, and if $v$ is idle then $v=\makespan(S)$. 
\end{description}


\begin{lemma}[frugality]\label{lem: frugal schedules}
Fix some $s,k,g$, and let $S$ be an
$(s,k)$-schedule that realizes $U_{s,k,g}$, that is 
$S$ has at most $g$ gaps and $\makespan(S) = U_{s,k,g}$.
Then $S$ is frugal.
\end{lemma}

\begin{proof}
If $S$ violates (f1) then we can extend $S$ as follows. Let
$t =\makespan(S)$ and $w > t$ be the smallest time such that
\begin{equation} 					\label{eq:frugal-u}
  w \geq t + \load_k(t,w).
\end{equation}
(Recall that $\load_k(t,w) = \sum_{j\le k,\: t \le r_{j} < w} p_{j}$.)
This time $w$ can be found simply by setting initially $w=t+1$, and
iteratively replacing $w$ by the right-hand side of
(\ref{eq:frugal-u}). Note that for this time $w$ we have in fact
equality in (\ref{eq:frugal-u}). We can extend $S$ by the time interval $[t,w)$ 
in which we schedule all jobs $j<k$ with $t \le r_{j}<w$, according to the
earliest-deadline property.  The result is an $(s,k)$-schedule with at
most $g$ gaps, contradicting the maximality of $S$.

Now assume that $S$ satisfies (f1) but not (f2). Let $[u,v)$ be some
execution interval of job $k$ in $S$. If $S$ is idle at time $u-1$,
then we can move one unit of job $k$ from $u$ to
$t =\makespan(S)<d_{k}$.  If $v <t$ and $S$ is idle at $v$, then we can
proceed in the same manner, moving one unit of job $k$ from $v-1$ to
$t$.  In both cases, by (f1), we obtain an $(s,k)$-schedule. This
schedule has at most $g$ gaps and completion time $t+1$, contradicting
the maximality of $S$.
\end{proof}


\paragraph{Function $U_{s,k,g}(p)$.}
Now we extend the definition of $U_{s,k,g}$ as follows.  First, for any
integer $p \ge 0$, we define an \emph{$(s,k,p)$-schedule} as an
$(s,k)$-schedule for the modified instance where we change the release
time of $k$ to $\max\braced{r_s,r_k}$ and the
processing time of $k$ to $p$, that is $r_k\assign \max\braced{r_s,r_k}$
and $p_{k}\assign p$. (All jobs other than $k$ remain unchanged.) 
For $p=0$, the notion of an $(s,k,0)$-schedule is equivalent to
an $(s,k-1)$-schedule.
Let $1\le s \le n$, $0\le k \le n$, $0\le g \le n-1$ and $p\ge 0$. 
We then define $U_{s,k,g}(p)$ as 
the maximum completion time of an $(s,k,p)$-schedule with at most $g$
gaps.  Naturally, for $p=0$, we have $U_{s,k,g}(0) = U_{s,k-1,g}$.

The idea behind the definition above is quite simple. Let $S$ be an
$(s',k)$-schedule, and $[r_s,t)$ be an interval such that the jobs
$j<k$ scheduled by $S$ in $[r_s,t)$ are exactly the jobs $j<k$
released in the same interval. Assume in addition that all these jobs
complete not later than $t$. Then the portion of $S$ in $[r_s,t)$ is
an $(s,k,p)$-schedule, where $p$ is the amount of job $k$ scheduled by
$S$ in this interval. The reason for adjusting $r_k$ is that we want
to allow $(s,k,p)$-schedules to schedule a portion of job $k$ even
if $r_k < r_s$. By changing $r_k$ to $r_s$ in this case, we include $k$
among the jobs that can be scheduled.

The following lemma will be useful in the proof of correctness of our
algorithm.


\begin{lemma}[expansion] \label{lem:expansion} Fix any $s,k,g$ and
  $p<p_{k}$ such that $r_k \le U_{s,k,g}(p)$. If $U_{s,k,g}(p)<d_k$, then
  $U_{s,k,g}(p+1)>U_{s,k,g}(p)$ and if $U_{s,k,g}(p)=d_k$, then
  $U_{s,k,g}(p+1)=d_k$ as well.
\end{lemma}


\begin{proof} Let $S$ be a schedule that realizes $U_{s,k,g}(p)$.
We examine the two cases in the lemma.

Consider first the case $\makespan(S) < d_k$. For $p>0$ we
argue as follows.
By Lemma~\ref{lem: frugal schedules} we know that $S$ is frugal,
so no job $j\le k$ is released at time $\makespan(S)$. Thus
appending one unit of job $k$ at
$\makespan(S)$ produces an $(s,k)$-schedule with at most $g$ gaps
and larger completion time. Therefore $U_{s,k,g}(p+1)>U_{s,k,g}(p)$.
For $p=0$ the argument is the same, with the only difference being
that we apply Lemma~\ref{lem: frugal schedules} to $k-1$ instead of $k$. 
In this case, among jobs $j\le k$ only job $k$ may be released at time 
$\makespan(S)$, so we can still append on unit of $k$ to $S$.

Now consider the case $\makespan(S)=d_{k}$ and let $[u,d_{k})$ be
the last block of $S$. We extend the support of $S$ by the time unit
$[u-1,u)$. Set $p_{k}\assign p+1$ and schedule jobs using the
earliest-deadline rule inside this new support. This new schedule
$S'$ will be identical to $S$ in $[r_s,u-1)$.

First we claim that in $S'$ the slot $u-1$ will not remain idle. 
Indeed, otherwise we would have that all jobs scheduled in $[u,d_{k})$
are released in that interval. These jobs include
job $k$ whose one unit is scheduled at $d_k-1$, by the
assumption about different deadlines. Since $p<p_{k}$, this would
contradict the feasibility assumption~(\ref{feasibility}) for the interval $[u,d_{k})$.  
(Note that the job scheduled at $u-1$ is not necessarily job $k$.)  Second,
in this new schedule no job will complete later than in $S$, so all
deadlines are met. This shows that $U_{s,k,g}(p+1)=d_{k}$, as claimed.
\end{proof}


\paragraph{Schedule compression.}
In the previous section, in the proof of the partitioning lemma, at
one point we were gradually compressing a unit-jobs schedule.  We
generalize this operation now to arbitrary-length jobs.

Fix any $s,k',p$. (We use notation $k'$ now instead of $k$, to
avoid confusion later in this section where
the results derived below will be used with either $k' = k-1$
or $k' = k$. Also later, in Section~\ref{sec: PanyLany}, we will use $k' = n$.) 
Let $T$ be some $(s,k')$-schedule and
$[w,v)$ the last block in $T$, where $v = \makespan(T)$.
The compression of $T$ consists of
reducing its completion time, without increasing the number of gaps.
It is accomplished by applying one of the steps below,
{\Truncate} or {\ShiftBack}, depending on whether the slot $v-1$ of $T$
is fixed or not. We remark here that
the resulting schedule may not be frugal. 

\begin{description}
	
\item{{\Truncate}:} Suppose that slot $v-1$ is fixed, and let
$[r_{i},v)$ be the fixed segment containing $v-1$, with maximal $r_i$.
The job $i$ can be found by a simple procedure:  Initially, let $i$ be
the job scheduled at $v-1$.  Then iteratively replace $i$ with the
job $j$ scheduled in $[r_i,v)$ that minimizes $r_j$, until a fixed point
is reached.  

Now, remove $[r_{i},v)$ from $T$ and let $T'$
be the resulting schedule.  By definition of fixed segments, all jobs
scheduled in $[r_i,v)$ are released in this segment.  Therefore $T'$
is an $(s,k')$-schedule, and if $r_{i}-1$ is idle (and $i\neq s$), $T'$
has one gap less than $T$, otherwise the number of gaps remains the
same. By the definition of fixed schedules, $T'$ schedules all jobs
of $T$ that are released before $r_i$.

\item{{\ShiftBack}:} Suppose that slot $v-1$ of $T$ is not fixed.  In
  this case we modify $T$ as follows: For each non-fixed slot in
  $[w,v)$, move the job unit in this slot to the previous non-fixed
  slot. The job unit scheduled in the first non-fixed slot in this
  block will move to $w-1$. Let $T'$ be the resulting schedule.

  Note that if $t$, $w\le t < v$, is a non-fixed slot executing some
  job $i$ and $t' < t$ is the previous non-fixed slot (that is, all
  slots between $t'+1$ and $t$ are fixed), then, by the definition of
  fixed slots, we have $r_i \le t'$. Therefore shifting the schedule,
  as above, will not violate release times, and we conclude that $T'$
  is an $(s,k')$-schedule with $\makespan(T') = \makespan(T) -1$.  If
  $w-2$ is not idle, $T'$ has one gap less than $T$, otherwise the
  number of gaps remains the same. Also, $T'$ schedules all jobs of $T$.
\end{description}

Both operations, {\Truncate} and {\ShiftBack}, convert $T$ into
another $(s,k')$-schedule $T'$
with $\makespan(T') < \makespan(T)$, and with the number of gaps in $T'$
not exceeding the number of gaps in $T$.
In what follows, we will also use the fact that
{\ShiftBack} reduces the completion time only by $1$.


\begin{lemma}[compression lemma]              \label{lem: compression} 
Fix any $s$, $k'$, and consider a time step $\theta \ge r_s$
that satisfies the following condition: for each job $j\le k'$, 
if $r_s\le r_j < \theta$ then $\greedyC_{s,j}\leq \theta$.
Suppose that there is an $(s,k')$-schedule $Q$ with completion time
$\makespan(Q)>\theta$ and at most $g$ gaps. Then there is an 
$(s,k')$-schedule $R$ that
schedules all jobs $j\le k'$ with $r_s\le r_j < \theta$ and satisfies
the following properties:
\begin{description}
\item{\mbox{\rm (a)}}
$\makespan(R)\leq \theta$ and the number of
gaps in $R$ is at most $g$, and

\item{\mbox{\rm (b)}} 
if $\makespan(R) < \theta$ then the number of gaps in $R$
is strictly less than $g$.
\end{description}
\end{lemma}

\begin{proof}
Starting from $Q$, we repeatedly apply
  the compression steps {\Truncate} and {\ShiftBack} described above,
  until we obtain a schedule $R$ with $\makespan(R)\le \theta$.  As
  explained above, the compression steps do not increase the number of
  gaps and $R$ schedules all jobs of $Q$ released before $\theta$.
Thus (a) holds.

To prove (b), suppose $\makespan(R) < \theta$. Since {\ShiftBack} reduces
the completion time by $1$ only, this is possible only if the
compression process ended with a  {\Truncate} step.
Denote by $T$ the schedule right before this step and let $[r_i,v)$
be the fixed segment truncated from $T$ in this step, where
$\makespan(T) = v > \theta$. 

If $r_i \ge \theta$ then, since $\makespan(R) < \theta$, $T$ had a gap
$[\makespan(R),r_i)$ that will be eliminated in the last step.
So the number of gaps in $R$ is strictly less than $g$.

Thus, to complete the proof, it is sufficient to show that we must
have $r_i\ge \theta$. Towards contradiction, suppose that $r_i < \theta$.
All slots of $T$ in $[\theta,v)$ are fixed, so, by the assumptions
of the lemma and by Lemma~\ref{lem:fixed=earliest}, they cannot contain
any jobs released before $\theta$. But then the choice of $r_i$ in
procedure {\Truncate} implies that $r_i < \theta$ is not possible, as claimed.
\end{proof}


\paragraph{Function $P_{s,k,g,l}$.}
We now extend somewhat the notion of gaps. Let $S$ be
an $(s,k)$-schedule and $t \ge \makespan(S)$.
A \emph{gap of $S$ with respect to $[r_s,t)$} is either
a gap of $S$ (as defined before) or the interval
$[\makespan(S),t)$, if $\makespan(S)< t$.

\begin{figure}[ht]
\centerline{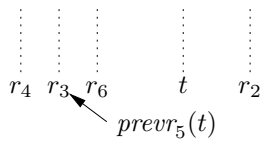}
\caption{Illustration of the definition of $\prevr_{k'}(t)$.}
\label{fig:prevr}
\end{figure}

For any job $k'$ and time $t$, let $\prevr_{k'}(t)$ be the latest release 
of a job $j \le k'$ before $t$, that is
\begin{eqnarray*}
	\prevr_{k'}(t) &=& \max\braced{ r_j \suchthat j \le k' \,\&\, r_j < t}.
\end{eqnarray*}
If there is no such job $j$, we take $\prevr_{k'}(t) = -\infty$.
(See Figure~\ref{fig:prevr} for illustration.)


\begin{figure}[ht]
\centerline{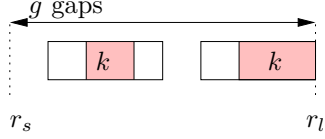}
\caption{Roughly (but not exactly),
$P_{s,k,g,l}$ is the minimum value $p$ such that the modified instance with
$r_k\assign\max\braced{r_k,r_s}$ and $p_k\assign p$ has
an $(s,k)$-schedule with at most $g$ gaps and completion time $r_l$.}
\label{fig:Pskgl}
\end{figure}

We define another table $P_{s,k,g,l}$, where the indices range
over all $s=1,\ldots,n$,  $k = 1,\ldots,n$, $g=0,\ldots,n-1$ and
$l=1,\ldots,k-1$ for which $r_l \ge r_s$.
$P_{s,k,g,l}$ is the minimum amount $p\geq 0$ 
of job $k$ for which there is an $(s,k,p)$-schedule $S$ that
satisfies $\prevr_{k-1}(r_l) < \makespan(S) \le r_l$
and has at most $g$ gaps with respect to $[r_s,r_l)$. (See Figure~\ref{fig:Pskgl}.)
By convention, $P_{s,k,g,l}=+\infty$ if there is no such $p$.
In particular, for $r_l = r_s$ (which is equivalent to $l=s$, so it is possible only
for $s\le k$) we have $P_{s,k,g,s}= 0$, and this value is
realized by the empty $(s,k)$-schedule.
Note also that for $r_k \geq r_l$, the value of $P_{s,k,g,l}$ is either $0$ or
$+\infty$, depending on whether there exists or not
an $(s,k-1)$-schedule $S$ that satisfies the condition above.

\begin{lemma}[extremal values of $P$] \label{lem:P=0}
{\mbox{\rm (a)\,}}
If there is a job $j<k$ released in $[r_s,r_l)$ with
    $\greedyC_{s,j}>r_l$, then $P_{s,k,g,l}=+\infty$. 

{\mbox{\rm (b)}\,}
 $P_{s,k,g,l}=0$ if and only if $U_{s,k-1,g}\geq r_l$ and every
  job $j<k$ released in $[r_s,r_l)$ satisfies $\greedyC_{s,j}\le r_l$.
\end{lemma}

\begin{proof}
To show (a), suppose that for some (finite) $p$ there is an
$(s,k,p)$-schedule $S$ with $P_{s,k,g,l} = p$.
Then, by the definition of $P_{s,k,g,l}$,
every job $j\leq k$ released in $[r_s,r_l)$ is scheduled by $S$ and
therefore $\greedyC_{s,j}\leq r_l$.

We now show (b).
Suppose that $P_{s,k,g,l} = 0$. By part (a), every
job $j<k$ released in $[r_s,r_l)$ satisfies $\greedyC_{s,j}\le r_l$.
Let $S$ be an $(s,k-1)$-schedule that realizes $P_{s,k,g,l}$.
In particular, $S$ schedules all jobs $j < k$ released in $[r_s,r_l)$. 
Let $T$ be the $(l,k-1)$-schedule
with completion time $U_{l,k-1,0}$ and no gaps. Note that $T$ is not
empty, since it schedules $l$. Then the union of $S$
and $T$ is an $(s,k-1)$-schedule with at most $g$ gaps and
completion time at least $r_l+1$, which shows $U_{s,k-1,g} > r_l$.

To show the reverse implication, assume that
$U_{s,k-1,g}\geq r_l$ and that every
job $j<k$ released in $[r_s,r_l)$ satisfies $\greedyC_{s,j}\le r_l$.
Let $S$ be an $(s,k-1)$ schedule that realizes $U_{s,k-1,g}$, that is,
$S$ has at most $g$ gaps and completion time $U_{s,k-1,g}\geq r_l$.  
If we have equality we are done. Otherwise, $S$ satisfies the
assumptions of the compression lemma,
Lemma~\ref{lem: compression} (with $k' = k-1$ and $\theta = r_l$). 
By applying this lemma, we obtain an $(s,k-1)$-schedule $R$ with 
$\makespan(R) \le r_l$. The conditions (a) and (b) of
Lemma~\ref{lem: compression} imply that $R$ has at most $g$ gaps
with respect to $[r_s,r_l)$.
\end{proof}


Intuitively, an execution interval is \emph{internal} if its removal
creates a gap. For a formal definition, let $S$ be an $(s,k,p)$-schedule. 
An execution interval $[u,v)$ of job $k$ in
$S$ is called an \emph{internal execution interval of $k$}
if (i) $v$ is not idle and (ii) $u-1$ is not idle or $u=r_s$.
By extension, if $\makespan(S)\le t$,
we call $[u,v)$ an \emph{internal execution interval of $k$
with respect to $[r_s,t)$} if 
(i) $v$ is not idle or $v=t$, and (ii) $u-1$ is not idle or $u=r_s$. 


\begin{lemma}[internal execution intervals]               \label{lem:P}
  Let $p=P_{s,k,g,l}$ and assume $p<+\infty$.  Let $S$ be an
  $(s,k,p)$-schedule that realizes $P_{s,k,g,l}$. Then
  \begin{description}
  \item{{\mbox{\rm (a)}\,}}
    Every execution interval of $k$ in $S$ is an internal execution
	interval with respect to $[r_s,r_l)$.  Moreover, if $p>0$ then $S$
	contains exactly $g$ gaps with respect to $[r_s,r_l)$.
      \item{{\mbox{\rm (b)}\,}}
	Let $[u,t)$ be some execution interval of $k$, $h$ be the
	  number of gaps before $u$ in $S$, and $q$ the amount of $k$
	  scheduled in $[r_s,u)$ by $S$.  Then $u=U_{s,k,h}(q)$.
  \end{description}
\end{lemma}

\begin{proof}
Part (a) of the lemma follows simply from the minimality of $p$. If $S$
had a non-internal execution interval of $k$, we can remove this
interval, reducing $p$, without increasing the number of gaps.
Similarly, if the number of gaps is less than $g$, we can remove
any execution interval of $k$. 

We now show part (b). By (a), $[u,t)$ is an internal execution of $k$ with
respect to $[r_s,r_l)$.
By the earliest deadline property, all jobs $j < k$ with $r_s\le r_j < u$
are completed before $u$. So the segment of $S$ between $r_s$ and $u$ is
an $(s,k,q)$-schedule with $h$ gaps and completion time $u$
(because either $u = r_s$ or slot $u-1$  is not idle), so $U_{s,k,h}(q)\geq u$.

If $U_{s,k,h}(q) = u$ we are done. Thus it remains to show that
$U_{s,k,h}(q) > u$ is impossible. Towards contradiction, 
assume $U_{s,k,h}(q) = u' >u$ and let $Q$ be an $(s,k,q)$-schedule with at 
most $h$ gaps and completion time $u'$.

We have two cases. If $u' \le t$, consider schedule $S'$ which is
the union of $Q$ and the portion of $S$ between $u'$ and $r_l$.
Denoting $p' = p + u - u'$, we get that
$S'$ is an $(s,k,p')$-schedule with at most $g$ gaps with respect to $[r_s,r_l)$.
Since $p' < p$, this contradicts the definition of $S$.

Now, suppose that $u' > t$. We apply Lemma~\ref{lem: compression} to $Q$,
with $k' = k$ and $\theta = t$, to obtain
a contradiction similar to the previous case.
To verify that the assumptions of Lemma~\ref{lem: compression} hold,
consider the modified instance where
$r_k\assign \max\braced{r_k,r_s}$ and $p_k \assign q$.  
For this modified instance, $\greedyC_{s,k}\leq u < \theta$.
Also, by the earliest deadline property, every job $j<k$ released in
$[r_s,t)$ completes not later than at $u$ in $S$ (in particular,
no job $j < k$ is released in $[u,t)$). Therefore $\greedyC_{s,j}\leq u < \theta$.  

The compression lemma gives us
an $(s,k,q)$-schedule $R$ scheduling all jobs $j \le k$
released in $[r_s,u)$ that satisfies condition (a) and (b)
of that lemma, with $\theta = t$. Consider schedule $S''$ which is
the union of $R$ and the portion of $S$ between $t$ and $r_l$.
(Note that, unlike in the previous case, the slots between
$\makespan(R)$ and $t$ are left idle.)
$S''$ is an $(s,k,p'')$ schedule with $p'' = p + u - t < p$.
We now have two sub-cases, depending on whether 
$\makespan(R) = t$ or $\makespan(R) < t$. In both sub-cases
though, using properties (a) and (b) from Lemma~\ref{lem: compression},
we conclude that $S''$ has at most $g$ gaps, which, together
with $p'' < p$, contradicts the definition of $S$.
\end{proof}


\paragraph{Outline of the algorithm.}
The algorithm in this section computes both functions $U_{s,k,g}$ and
$P_{s,k,g,l}$.  The intuition is this. Let $S$ be an $(s,k)$-schedule
that realizes $U_{s,k,g}$, that is $S$ has
at most $g$ gaps and completion time $u = \makespan(S) = U_{s,k,g}$.
If $S$ does not schedule $k$ then $u = U_{s,k-1,g}$. 

\begin{figure}[t]
\centerline{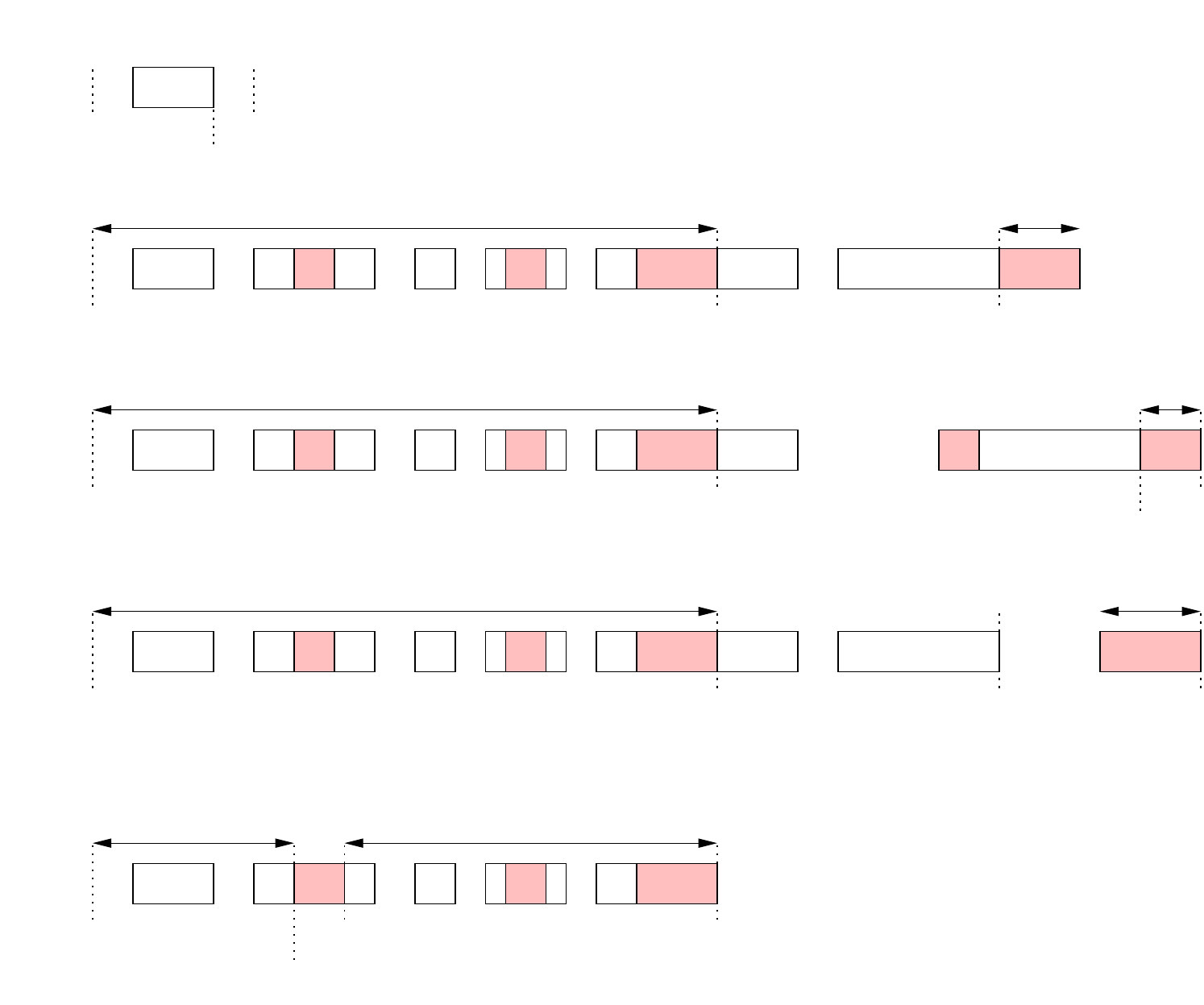}
\caption{Illustration of recurrence equations in Algorithm~{\algB}.}
\label{fig: correctness B}
\end{figure}

So assume that $S$ schedules job $k$. There are several cases.
Consider, for example, the case when $u<d_{k}$ and when $k$ has an
execution interval $[t',t)$ with $t < u$.  (See the second case in
Figure~\ref{fig: correctness B}.) Take $[t',t)$ to be the last such
interval. Since $S$ is frugal, we know that $S$ is not idle at $t'-1$
and at $t$. Then, by the earliest-deadline property, $S$ schedules at
$t$ some job $l<k$ with $r_l=t$.  Now, the part of $S$ up to $r_{l}$
has some number of gaps, say $h$. The key idea is that, roughly,
the amount $q$
of job $k$ in this part is minimal among all $(s,k,q)$-schedules with
completion time $r_l$ and at most $h$ gaps, so this amount is
equal to $P_{s,k,h,l}$.  Otherwise, if it were not minimal, then we
could replace the part of $S$ before $t$ by an $(s,k,q')$-schedule for
some $q'<q$ and this would imply $U_{s,k,g}(p)\geq U_{s,k,g}(p_k)$ for
$p = p_k+q'-q <p_k$, contradicting Lemma~\ref{lem:expansion}.  By the
choice of $[t',t)$, and induction, the interval $[t,u)$ of $S$
consists of an $(l,k-1)$-schedule with at most $g-h$ gaps followed by
$p_k - P_{s,k,h,l}$ units of $k$, and thus $U_{s,k,g}$ can be
expressed as $U_{l,k-1,g-h} + p_k - P_{s,k,h,l}$.

If $u < d_k$ and $k$ has just one execution segment ending at $u$,
then there is no segment $[t',t)$ considered above. But then the
formula $U_{l,k-1,g-h} + p_k - P_{s,k,h,l}$ applies as well, since we
can take $l=s$ and $h=0$, and then $P_{s,k,h,l} = 0$, so in this case
$U_{s,k,g}$ will be equal to $U_{s,k-1,g}+p_k$.

The remaining case, when $u = d_k$, breaks into two sub-cases
depending on whether the last block contains only units of $k$ or
not. In order to determine whether it is possible to achieve
$u = d_k$ with only $g$ gaps, we proceed in a similar manner, by partitioning 
the schedule using an execution interval $[t',t)$ of $k$ (if it exists).

The idea behind the recurrence for $P_{s,k,g,l}$ is similar --
essentially, it consists of partitioning the schedule realizing $P_{s,k,g,l}$
into disjoint sub-schedules, with the first one ending at a release time 
of some job $j$.


\paragraph{Algorithm~{\algB}.}
The algorithm computes the values of
$U_{s,k,g}$ and $P_{s,k,g,l}$ in order of increasing $k$
and stores these values in tables $\UU{s}{k}{g}$ and $\PP{s}{k}{g}{l}$.

First, for $k=0$, we initialize
$\UU{s}{0}{g} \assign r_s$ for all $s= 1,...,n$ and $g= 0,...,n-1$.
Then, for $k = 1,...,n$ we do the following:
\begin{itemize}
	\item Compute $\PP{s}{k}{g}{l}$ for all $s = 1,...,n$,
	 	$g = 0,...,n-1$, and for $l = 1,...,k-1$ such that
		$r_l\ge r_s$. The indices $l$ are processed in order
		of increasing $r_l$.
	\item Compute $\UU{s}{k}{g}$ for all $s = 1,...,n$ and $g = 0,...,n-1$.
\end{itemize}
For $k\ge 1$, the values of $\PP{s}{k}{g}{l}$ and $\UU{s}{k}{g}$ are computed
using the recurrence equations described below. These equations are
illustrated in Figure~\ref{fig: correctness B}.
Once all these values are computed, the algorithm
determines the minimum number of gaps as the smallest $g$ for which
$\UU{1}{n}{g} > \max_j r_j$. (Recall that job $1$ is a special job of unit
length with minimum release time.)


\noindent
\emph{Computing $\PP{s}{k}{g}{l}$.}  
If there is a job $j < k$ such that $r_s\le r_j < r_l$ 
and  $\greedyC_{s,j} > r_l$, then $\PP{s}{k}{g}{l} \assign +\infty$.
Otherwise, we have that every job $j<k$ such that $r_s\le r_j < r_l$ 
satisfies $\greedyC_{s,j}\leq r_l$.

If $\UU{s}{k-1}{g} \geq r_l$ then $\PP{s}{k}{g}{l} \assign 0$. (Note
that this will include the case $s = l$, if $s\le k$.) 
In the remaining case, we have 
$\UU{s}{k-1}{g} < r_l$; thus in particular also $r_s < r_l$.
We then compute $\PP{s}{k}{g}{l}$ recursively as follows:
\begin{eqnarray}   
\PP{s}{k}{g}{l} \;\assign\;   
\min_{\begin{subarray}{c} 0\le h\le g \\ j<k\\ r_s< r_j \le r_l \end{subarray}} 
\braced{ r_{j}-\UU{s}{k-1}{h}+\PP{j}{k}{g-h}{l}
			 \suchthat 
			 \prevr_{k-1}(r_j) <  \UU{s}{k-1}{h}  < r_j
				\;\&\; r_k \le \UU{s}{k-1}{h}
   }
   \label{eqn: recurrence for P}
\end{eqnarray}
As usual, by default, if the conditions in the minimum are not
satisfied by any $h,j$, then $\PP{s}{k}{g}{l}$ is assumed to be $+\infty$. 


\noindent
\emph{Computing $\UU{s}{k}{g}$.}  $\UU{s}{k}{g}$ is computed
recursively as follows.  If $r_k < r_s$ then we let $\UU{s}{k}{g}
\assign \UU{s}{k-1}{g}$. Otherwise, for $r_k \geq r_s$, we let
\begin{align}
\UU{s}{k}{g} &\assign 
   \max_{l,h}
	\left\{\begin{array}{ll}
	\UU{s}{k-1}{g} &\textrm{if } \UU{s}{k-1}{g} < r_k
\\[0.7em]
       \UU{l}{k-1}{g-h}+p_{k}-\PP{s}{k}{h}{l}
                & \textrm{if }\PP{s}{k}{h}{l}\leq p_{k},
		\\
		& d_{k}- \UU{l}{k-1}{g-h} > p_{k} - \PP{s}{k}{h}{l},
                  \\
                  &\UU{l}{k-1}{g-h}\ge r_k
                  \textrm{ and } 
                \\
                & \UU{l}{k-1}{g-h} >\prevr_{k-1}( \UU{l}{k-1}{g-h}+p_{k}-\PP{s}{k}{h}{l} )
\\[0.7em]
       d_k      &\textrm{if }\PP{s}{k}{h}{l}<p_{k},
		\\
 		& d_{k}- \UU{l}{k-1}{g-h} \le p_{k} - \PP{s}{k}{h}{l} \textrm{ and}
                \\
                & \UU{l}{k-1}{g-h} > \prevr_{k-1}(d_k)
\\[0.7em]
       d_{k}	&\textrm{if }h<g, \PP{s}{k}{h}{l}<p_{k},
		\\
		& d_{k}- \UU{l}{k-1}{g-h-1} > p_{k} - \PP{s}{k}{h}{l}  \textrm{ and}
		\\
 		& \UU{l}{k-1}{g-h-1} > \prevr_{k-1}(d_k)
           \end{array}
          \right.
                \label{eqn: recurrence for U}
\end{align}
In this formula,
the maximization ranges over all pairs $l,h$ where $1\leq l<k, 0\leq h\leq g$,
and for $s>k$ we include one additional pair $l=s,h=0$, for which
the value of $\PP{s}{k}{0}{s}$ should be interpreted as $0$. (Recall that
$P_{s,k,0,s}$ is not defined for $s > k$.)


\begin{lemma}[correctness of {\algB}]\label{lem: correctness B}
Algorithm~{\algB} correctly computes the values of $U_{s,k,g}$ and
$P_{s,k,g,l}$.  More specifically, for all $s =
1,...,n$, $k = 0,...,n$, and $g = 0,...,n-1$ we have $\UU{s}{k}{g} = U_{s,k,g}$ and
$\PP{s}{k}{g}{l} = P_{s,k,g,l}$ for $k > 0$ and all $l = 1,...,k-1$.
\end{lemma}

\begin{proof}
We show that there are schedules that
realize the values $\UU{s}{k}{g}$ and $\PP{s}{k}{g}{l}$ 
(the feasibility condition) and that these
values are indeed optimal. More specifically, 
we prove the following four properties.


\begin{description}

\item{\emph{Feasibility of $\PP{s}{k}{g}{l}$:\/}} For each $s$, $k$, 
  $g$ and $l$ for which $\PP{s}{k}{g}{l}  = p < +\infty$ there is an
  $(s,k,p)$-schedule $T_{s,k,g,l}$ with  
$\prevr_{k-1}(r_l) < \makespan(T_{s,k,g,l}) \le r_l$
 and at most $g$ gaps with respect to $[r_s,r_l)$.

\item{\emph{Optimality of $\PP{s}{k}{g}{l}$:\/}} 
  $\PP{s}{k}{g}{l} \le P_{s,k,g,l}$, for all $s$,  $k$, $g$ and $l$.

\item{\emph{Feasibility of $\UU{s}{k}{g}$:\/}} For each $s$, $k$ and
$g$, if $\UU{s}{k}{g}$ is defined then
there is an $(s,k)$-schedule $S_{s,k,g}$ with completion time
$\UU{s}{k}{g}$ and at most $g$ gaps.
	
\item{\emph{Optimality of $\UU{s}{k}{g}$:\/}} 
 $\UU{s}{k}{g} \ge U_{s,k,g}$, for all $s$, $k$ and $g$.
\end{description}

Note that the last condition implies that $\UU{s}{k}{g}$ is 
always defined; therefore the feasibility condition for
$\UU{s}{k}{g}$ applies, in fact, to all values of $s$, $k$, $g$
in the appropriate range. A similar comment applies to
$\PP{s}{k}{g}{l}$, although in this case $\PP{s}{k}{g}{l}$
is defined only when $P_{s,k,g,l}$ is.

The proof is by induction on $k$. Consider first $k=0$.
In this case we only need to prove the feasibility and
optimality of $\UU{s}{0}{g}$ (since $P_{s,k,g,l}$ and $\PP{s}{k}{g}{l}$
are not defined for $k=0$). We take $S_{s,0,g}$ to be
the empty schedule, which is trivially feasible and has completion
time $r_s=\UU{s}{0}{g}$. On the other hand, there is only one
$(s,0)$-schedule, namely the empty schedule, which has completion time
$r_{s}$, proving the optimality of $\UU{s}{0}{g}$.

Now fix some $s, k, g,l$ where $k\ge 1$, $l < k$ and $r_l\ge r_s$. 
Assume that the feasibility and optimality condition for
$\UU{s'}{k-1}{g'}$ is true for any $s'$, $g'$. We show the
feasibility and optimality of $\PP{s}{k}{g}{l}$.


\emparagraph{Feasibility of $\PP{s}{k}{g}{l}$:\/} 
We assume that $\PP{s}{k}{g}{l}$ is finite and we prove the existence
of $T_{s,k,g,l}$ by induction on $r_l - r_s$. If $r_l = r_s$ then 
$\PP{l}{k}{g}{l} =  P_{l,k,g,l} = 0$, and we take $T_{s,k,g,s}$ to be the 
empty $(s,k)$-schedule. 

So assume now that $r_s < r_l$. We can also assume that 
every job $j<k$ released in $[r_s,r_l)$ satisfies $\greedyC_{s,j}\le r_l$
(for otherwise $\PP{s}{k}{g}{l} = +\infty$). We now distinguish two cases.

\noindent\mycase{1}
$\UU{s}{k-1}{g}\geq r_l$. 
By the algorithm, $\PP{s}{k}{g}{l} = 0$.
By induction, $U_{s,k-1,g}\ge r_l$ as well, and using
Lemma~\ref{lem:P=0} we get $P_{s,k,g,l} = 0$; in other words, there is
an $(s,k-1)$-schedule $T$ with at most $g$ gaps with respect to $[r_s,r_l)$.
This schedule $T$ can be constructed from $S_{s,k-1,g}$ by compression, as
described in the proof of Lemma~\ref{lem:P=0}.
Thus in this case we can take  $T_{s,k,g,l} = T$.

\noindent\mycase{2}
$\UU{s}{k-1}{g}< r_l$.
In this case, the algorithm will
compute  $\PP{s}{k}{g}{l}$ using recurrence (\ref{eqn: recurrence for P}).
Let $h,j$ be the values that realize the minimum  in (\ref{eqn: recurrence for P})
and denote $u = U_{s,k-1,h}$. Then
$\PP{j}{k}{g-h}{l}$ is finite, $p = r_j - u + \PP{j}{k}{g-h}{l}$,
$\prevr_{k-1}(r_j) < u < r_j$ and $r_k \le u$.
The first of those inequalities implies
that there are no jobs $i < k$ released in $[u,r_j)$.
We let $T_{s,k,g,l}$ be the union of
schedules $S_{s,k-1,h}$ and $T_{j,k,g-h,l}$ -- that both exist, by
induction -- with additional $r_j - u$ units of $k$ scheduled in the
interval $[u,r_j)$. (Note that we may have $j=l$, in which case
schedule $T_{j,k,g-h,l}$ will be empty.) Then $T_{s,k,g,l}$ is a feasible
$(s,k,p)$-schedule with at most $g$ gaps with respect to $[r_s,r_l)$.
Since $\makespan(T_{s,k,g,l}) = \makespan(T_{j,k,g-h,l})$, we also have 
$\prevr_{k-1}(r_l) < \makespan(T_{s,k,g,l}) \le r_l$, as required.


\emparagraph{Optimality of $\PP{s}{k}{g}{l}$:\/} 
The proof is by induction on
$r_{l}-r_{s}$.  For the base case $r_{s}=r_{l}$ we have
$\PP{s}{k}{g}{s}=0 \le P_{s,k,g,s}$.  Now assume $r_{l}>r_{s}$. 

We can assume $P_{s,k,g,l}<+\infty$, since otherwise $\PP{s}{k}{g}{l} \le P_{s,k,g,l}$
is trivial. Then, by Lemma~\ref{lem:P=0}(a), every job $j<k$ released in
$[r_s,r_l)$ satisfies $\greedyC_{s,j}\leq r_l$.  
If $\PP{s}{k}{g}{l} = 0$ then $\PP{s}{k}{g}{l} \le P_{s,k,g,l}$ is trivial again,
so we can assume that $\PP{s}{k}{g}{l} > 0$. By the algorithm, this
implies that  $\UU{s}{k-1}{g} < r_l$
(because the value of recurrence (\ref{eqn: recurrence for P}) cannot be $0$).
Therefore by Lemma~\ref{lem:P=0} we have $P_{s,k,g,l}>0$.

Let $T$ be a schedule that realizes $P_{s,k,g,l} = p$, that is
$T$ is an $(s,k,p)$-schedule with $\prevr_{k-1}(r_l) < \makespan(T) < r_l$
and at most $g$ gaps with respect to $[r_s,r_l)$. Let $[u,t)$ be the first
execution interval of $k$ in $T$ and $h$ the number of gaps before
$u$.  By Lemma~\ref{lem:P}(a), $[u,t)$ is an internal execution interval of $T$ with
respect to $[r_s,r_l)$, so there is a job $j < k$ with $r_j = t$. (We may have 
$t = r_l$, in which case, obviously, $j=l$.) By the minimality of
$p$, the segment of $T$ in $[r_j,r_l)$ schedules $P_{j,k,g-h,l}$
units of $k$ and, by the induction hypothesis, this equals $\PP{j}{k}{g-h}{l}$.
By Lemma~\ref{lem:P}(b) we have $u=U_{s,k-1,h}$ which by the
induction hypothesis equals $\UU{s}{k-1}{h}$.  
The earliest-deadline property applied to $T$ implies there is no job $i<k$ released in
$[u,r_j)$, that is $\prevr_{k-1}(r_j) < u < r_j$. Therefore $h,j$ are a
valid choice for the recurrence (\ref{eqn: recurrence for P}), 
and $\PP{s}{k}{g}{l} \leq p$ follows.


At this point we can assume the
feasibility and optimality conditions for $\PP{s'}{k}{g'}{l'}$ and
$\UU{s'}{k-1}{g'}$, for any $s'$, $l'$ and $g'$.
Thus, to streamline the arguments, in the rest of the proof we will interchangingly use
notations $\PP{s'}{k}{g'}{l'}$ and $P_{s,k,g,l}$, as well as
$\UU{s'}{k-1}{g'}$ and $U_{s,k-1,g}$, without an explicit reference
to the inductive assumption.  We show the
feasibility and optimality of $\UU{s}{k}{g}$.


\emparagraph{Feasibility of $\UU{s}{k}{g}$:\/} Here we will show how
we can construct $S_{s,k,g}$ using the recurrence for $\UU{s}{k}{g}$.
We consider cases corresponding to those in the algorithm.

Suppose first that $r_k < r_s$, in which case $\UU{s}{k}{g} = \UU{s}{k-1}{g}$.
In this case we take $S_{s,k,g} =
S_{s,k-1,g}$. By induction, $S_{s,k,g}$ is a feasible
$(s,k-1)$-schedule with completion time $\UU{s}{k}{g}$, and is also a 
feasible $(s,k)$-schedule, by the assumption about $r_k$.

Assume now that $r_k \ge r_s$. We now  construct $S_{s,k,g}$ for each of
the four cases in the maximum (\ref{eqn: recurrence for U}).

\noindent\mycase{1}
$\UU{s}{k}{g}$ is realized by the first option.
Then we set $S_{s,k,g}=S_{s,k-1,g}$, which by the case condition $r_k > \UU{s}{k-1}{g}$ 
is  an $(s,k)$-schedule with completion time $\UU{s}{k-1}{g}=\UU{s}{k}{g}$.

\noindent\mycase{2}
$\UU{s}{k}{g}$ is realized by the second option, for some values $l,h$.
Then let $u = \UU{l}{k-1}{g-h}$ and $t = \UU{s}{k}{g}$. 
By the case conditions, $r_k\le u\le t<d_k$. We define $S_{s,k,g}$ to be a union
of $T_{s,k,h,l}$ and $S_{l,k-1,g-h}$, whose existence follows from induction,
with additional $t-u$ units of $k$ scheduled in the interval $[u,t)$. 
(In the special case $l=s>k$ and $h=0$,
we take $T_{s,k,h,l}$ to be the empty schedule.)  By the case conditions, there
are no jobs $j<k$ released in $[u,t)$, so $S_{s,k,g}$ is a feasible
$(s,k)$-schedule with completion time $t$ and at most $g$ gaps.

\noindent\mycase{3}
$\UU{s}{k}{g}$ is realized by the third option, for some $l,h$. Let
$u=\UU{l}{k-1}{g-h}$, and $p=\PP{s}{k}{h}{l}+d_{k}-u$. We have
$u\le d_{k-1} <d_k$ and $p\le p_k$, by the case conditions.
Also, $r_k\leq d_k-p_k \le d_k - p \le u$. Now let $S$ be the union of the
$T_{s,k,h,l}$ and $S_{l,k-1,g-h}$, whose existence follows from induction,
followed by $d_{k}-u$ units of job $k$. (For $l=s>k$ and $h=0$,
we take $T_{s,k,h,l}$ to be the empty schedule.)
Then $S$ is an $(s,k,p)$-schedule with completion time $d_{k}$ and
at most $g$ gaps.  By Lemma~\ref{lem:expansion}, there is an
$(s,k)$-schedule $S_{s,k,g}$ (scheduling all $p_{k}$ units of job $k$)
with completion time $d_k$ and at most $g$ gaps.

\noindent\mycase{4}
$\UU{s}{k}{g}$ is realized by the last option, for some $l,h$. Then let
$p = p_k - \PP{s}{k}{h}{l}$. We have $r_k\leq d_k-p_k \leq d_k-p$.
Define $S_{s,k,g}$ to be the union of $T_{s,k,h,l}$ and $S_{l,k-1,g-h-1}$, with
additional $p$ units of $k$ scheduled in the interval $[d_{k}-p,d_k)$. 
(As in the previous cases, for $l=s>k$ and $h=0$,
we take $T_{s,k,h,l}$ to be the empty schedule.)
The union of $T_{s,k,h,l}$ and $S_{l,k-1,g-h-1}$
contains at most $g-1$ gaps, and it schedules $\PP{s}{k}{h}{l} <
p_{k}$ units of job $k$.  Scheduling the remaining $p$ units of $k$ in
$[d_{k}-p,d_k)$ will create one more gap.  Therefore $S_{s,k,g}$
is a feasible $(s,k)$-schedule with completion time $d_k$ and at most $g$ gaps.


\emparagraph{Optimality of $\UU{s}{k}{g}$:\/} 
Let $t = U_{s,k,g}$ and 
let $S$ be an $(s,k)$-schedule that realizes $U_{s,k,g}$, that is,
$S$ has at most $g$ gaps and completion time $t$. 
We need to show $\UU{s}{k}{g}\ge t$.

If $S$ does not schedule $k$, then $t =U_{s,k-1,g}$. 
This can happen if either $r_k< r_s$ or $t < r_k$. 
If $r_k < r_s$ then, by the algorithm and induction,
$\UU{s}{k}{g}= \UU{s}{k-1}{g} = U_{s,k-1,g} = t$.
Similarly, if $t < r_k$ then, by induction,
$\UU{s}{k-1}{g} = U_{s,k-1,g} > r_k$ and using the first option
of the algorithm we have
$\UU{s}{k}{g} \ge \UU{s}{k-1}{g} = U_{s,k-1,g} = t$.
So from now on we assume that $S$ schedules $k$.

Our objective now is to identify two numbers $h,l$ and show 
that we can find a corresponding decomposition of $S$ that would allow us
to apply one of the last three options in (\ref{eqn: recurrence for U}) and 
induction, 
yielding $t\le \UU{s}{k}{g}$. The proof is broken into several cases.

\noindent
\smallskip	
{\mycase{1}} $t < d_k$.
If $S$ has an internal execution interval of $k$,
let $[u,v)$ be the last internal execution interval of $k$ of $S$.
We let $l<k$ be the job released and scheduled at $v$
(this job $l$ exists by the definition of internal execution intervals
and the earliest-deadline property), and we let $h$
be the number of gaps of $S$ in the segment of $S$ in $[r_s,v)$.
In the other case, if $S$ does not have an internal execution interval of $k$, 
we choose $h=0$, $l=s$, and in the argument below we use $u = v = r_s$.

Let $q$ be the number of units of $k$ scheduled by $S$ in $[r_s,v)$.
The segment of $S$ in $[r_s,v)$ is an $(s,k,q)$-schedule with $h$
gaps with respect to $[r_s,v)$, thus $q\ge P_{s,k,h,l}$.

In fact, we claim that $q = P_{s,k,h,l}$. 
For suppose, towards contradiction, that $q > P_{s,k,h,l}$.
Let $Q$ be the schedule that realizes $P_{s,k,h,l}$. Then we could 
replace the segment of $S$ in $[r_s,v)$ by $Q$, reducing the number of units 
of $k$ in $S$, without changing the number of gaps and the completion time 
of $S$. But this contradicts Lemma~\ref{lem:expansion}, so we can conclude
that $q = P_{s,k,h,l}$, as claimed.

Let	$[z,t)$ be the execution interval of $k$ at the end of $S$. (This interval 
could be empty, that is we allow here $z=t$.) In this case ($t<d_k$), the last block 
contains jobs other than $k$. Thus the segment of $S$ in
$[r_l,z)$ is an $(l,k-1)$-schedule with at most $g-h$ gaps, so
$z \le U_{l,k-1,g-h}$.

We claim that, in fact, we have $z = U_{l,k-1,g-h}$. Indeed, towards
contradiction, suppose that $z < z' = U_{l,k-1,g-h}$. 
Let $R$ be an $(l,k-1)$ schedule with at most $g-h$ gaps
that realizes $U_{l,k-1,g-h}$. For $t=z$ we obtain an immediate
contradiction with the definition of $S$, since we could
replace the segment of $S$ in $[r_l,d_k)$ by $R$, obtaining
an $(s,k)$-schedule with at most $g$ gaps and completion time greater
than $t$. So we can assume now that $z < t$. Then
we can modify $S$ as follows: replace
the segment $[r_l,z')$ of $S$ by $R$, and if $z' < t$ then 
append to it a segment of $t-z'$ units of $k$. 
This produces a $(s,k,p')$-schedule, with $p' < p_k$,
at most $g$ gaps and completion time at least $t$, giving us
a contradiction with Lemma~\ref{lem:expansion}.
Thus we indeed have $z = U_{l,k-1,g-h}$.

Summarizing, we have $z = U_{l,k-1,g-h}$, $q = P_{s,k,h,l}\le p_k$,
$d_k - z > t - z = p_k - q$, $z\ge r_k$, and
$z > \prevr_{k-1}(t)$, for $t = U_{l,k-1,g-h}+p_k - q$.
Thus, by induction,
the second option in (\ref{eqn: recurrence for U}) will
apply, and we obtain $\UU{s}{k}{g}\ge U_{l,k-1,g-h} + p_k -q = t$.

\smallskip
\noindent	
{\mycase{2}} $t = d_k$. As in the previous case, we need to identify
appropriate values for $l$ and $h$. This is more challenging
than in the previous case because for $t=d_k$ the schedule $S$
that realizes $U_{s,k,g}$ may have ``slack", and thus arguments based on
the tightness of $S$ do not apply.

To get around this issue, for $p \ge 0$,
let $\UU{s}{k}{g}(p)$ be the value computed by the algorithm for the
modified instance where $p_{k} \assign p$.
We claim that if $\UU{s}{k}{g}(p) = d_k$ then $\UU{s}{k}{g}(p+1)= d_k$
as well. To justify this claim, note that
if $\UU{s}{k}{g}(p)$ is realized by option three, then
$\UU{s}{k}{g}(p+1)$ will also be realized by option three, so its value
remains $d_k$. If $\UU{s}{k}{g}(p)$ is realized by option four, then
$\UU{s}{k}{g}(p+1)$ will be realized either by option four or
option three (this uses the fact that $U_{l,k-1,g-h}\ge U_{l,k-1,g-h-1}$), 
and thus its value remains $d_k$ as well. Thus the claim holds.

Define $p^\ast\leq p_k$ to be the minimum amount of job $k$ for which
$U_{s,k,g}(p^\ast)=d_k$. By the previous claim it is sufficient to prove
that  $\UU{s}{k}{g}(p^\ast) = d_k$. Thus for the rest of the proof we
simply assume that $p_k = p^\ast$.

With this assumption, we choose $l$ and $h$ using a method
analogous to that in the previous case: let
$[u,v)$ be the last internal execution interval of $k$ of $S$,
$l<k$ be the job released and scheduled at $v$,
and $h$ be the number of gaps of $S$ in the segment of $S$ in $[r_s,v)$.
In the special case when
$S$ does not have an internal execution interval of $k$, 
we choose $h=0$, $l=s$ and $u = v = r_s$.

Let $q$ be the number of units of $k$ scheduled by $S$ in $[r_s,v)$.
The segment of $S$ in $[r_s,v)$ is an $(s,k,q)$-schedule with $h$
gaps with respect to $[r_s,v)$, thus $q\ge P_{s,k,h,l}$.
In fact, we must have $q = P_{s,k,h,l}$, for otherwise, if
$q > P_{s,k,h,l}$, we could replace
the segment of $S$ in $[r_s,v)$ by a schedule $Q$
that realizes $P_{s,k,h,l}$. The resulting schedule would have
the same number of gaps as $S$ and completion time $d_k$,
but fewer units of job $k$, 
so we get a contradiction with the choice of $q^\ast$.

We now have two sub-cases.

\begin{description}

\item{{\mycase{2.1}}} $k$ is not the only job in the last block.
	As in the previous case, let $[z,d_k)$ be the last execution interval of $k$ in $S$. 
	We have $z > \prevr_{k-1}(d_k)$. Since the segment of $S$ in $[r_l,z)$ is
	an $(l,k-1)$-schedule with completion time $z$ and at most $g-h$ gaps, 
	we also have $z \le U_{l,k-1,g-h}$. We can thus conclude that $q < p^\ast$,	
	$d_k - U_{l,k-1,g-h} \le p^\ast - q$ and $U_{l,k-1,g-h} > \prevr_{k-1}(d_k)$.
	(Recall that $q = P_{s,k,h,l}$.) Therefore, applying the inductive assumption, 
	we obtain that the third option in (\ref{eqn: recurrence for U}) applies,
	yielding $\UU{s}{k}{g} = d_k = t$.

\item{{\mycase{2.2}}} $k$ is the only job in the last block.
	The minimality of $p^\ast$ implies that $p^\ast = q+1$, that is
	the last block is $[d_k-1,d_k)$, since otherwise we could remove
	from the schedule the units of $k$ right before the last one.
	(Recall that $q = P_{s,k,h,l}$.)
	Let $[t',d_k-1)$ be the last gap in $S$. Then $\prevr_{k-1}(d_k) < t'$. Since
	the segment of $S$ in $[r_l,t')$
	is an $(l,k-1)$-schedule with at most $g-h-1$ gaps, we also have
	$t' \le U_{l,k-1,g-h-1}$, so $\prevr_{k-1}(d_k) <  U_{l,k-1,g-h-1}$.
	Obviously, $h < g$ and $P_{s,k,h,l} = q < p^\ast$.
	Applying induction,
	if $d_k - U_{l,k-1,g-h-1} > p^\ast-q$, option four in (\ref{eqn: recurrence for U})
	will apply.	Otherwise, $d_k - U_{l,k-1,g-h-1}\le p^\ast-q$, in which case
	option three will apply, because $U_{l,k-1,g-h}\ge U_{l,k-1,g-h-1}$.
	(In fact, in this particular case, we would have equality, since
	$d_k - U_{l,k-1,g-h-1} \le p^\ast-q \le 1$ implies $U_{l,k-1,g-h-1} = d_{k-1} = d_k-1$.)
	In both of these cases we obtain $\UU{s}{k}{g} = d_k = t$.

\end{description}

We have now proved that in all cases we obtain $t\le \UU{s}{k}{g}$,
completing the proof of optimality of $\UU{s}{k}{g}$, and the lemma.
\end{proof}



\begin{theorem}\label{thm: algorithm B}
Algorithm~{\algB} correctly computes the optimum solution for
$1|r_j;{\pmtn}; L=1|E$, and it can be implemented in time $O(n^5)$.
\end{theorem}

\begin{proof}
The correctness follows from Lemma~\ref{lem: correctness B}.
The running time analysis is similar to the analysis of
Algorithm~{\algA}.  The table $\UU{s}{k}{g}$ is computed in time
$O(n^{5})$ since there are $O(n^{3})$ variables and each requires
minimization over $O(n^{2})$ values.  The table $\PP{s}{k}{g}{l}$ has
size $O(n^4)$. For each entry $\PP{s}{k}{g}{l}$, the job $j$ in the
recurrence is uniquely determined by $h$ (if it exists at all), so the
minimization requires time $O(n)$. Thus the total running time is
$O(n^{5})$.
\end{proof}


\section{Minimizing the Energy}
\label{sec: PanyLany}


We now show how to solve the general problem of minimizing the energy
for an arbitrary given value $L$. This new algorithm consists of
computing the table $U_{s,k,g}$ (using either Algorithm~{\algA} or
{\algB}) and an $O(n^2\log n)$-time post-processing. Thus we can solve the
problem for unit jobs in time $O(n^4)$ and for arbitrary-length jobs
in time $O(n^5)$.

Recall that for this general cost model, the cost (energy) is defined as 
the sum, over all gaps, of the minimum between $L$ and the gap length.
Call a gap \emph{small} if its length is at most $L$ and 
\emph{large} otherwise.  The idea of the
algorithm is this: We show first that there is an optimal schedule
where the short gaps divide the instance into disjoint
sub-instances (in which all gaps are large).
For those sub-instances, the cost is simply the number
of gaps times $L$. To compute the overall cost, we add to this
quantity the total size of short gaps.

Given two schedules $S$, $S'$ of the input instance, we say that $S$
\emph{dominates} $S'$ if there is a time point $t$ such that the
supports of $S$ and $S'$ in the interval $(-\infty,t)$ are identical and
$S$ schedules a job at time $t$ while $S'$ is idle.  This relation
defines a total order on all schedules.  The correctness of the
algorithm relies on the following separation lemma.


\begin{lemma} \label{dominated_schedules}
There is an optimal schedule $S$ with the following property: For any
small gap $[u,t)$ of $S$ and job $j$, if $C_j(S)\ge t$ then $r_j\ge t$.
\end{lemma}

\begin{figure}[ht]
\centerline{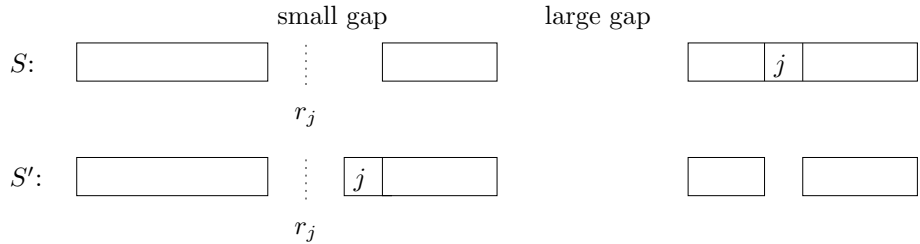}
\caption{Idea of the proof of Lemma~\ref{dominated_schedules}. Schedule $S'$ dominates $S$.}
\label{fig:idea-dom}
\end{figure}

\begin{proof}
Among all optimal schedules, choose $S$ to be one not dominated by another
optimal schedule, and let $[u,t)$ be a small gap in $S$ (see Figure~\ref{fig:idea-dom}).
If there is a job $j$ with $r_j < t$ such that a unit of $j$ is
scheduled at some time $t' \geq t$, then we can move this execution
unit to the time unit $t-1$. This will not increase the overall cost,
since the cost of the small gap decreases by one, and the idle time
unit created at $t'$ increases the cost at most by $1$.  The resulting
schedule, however, dominates $S$ -- contradiction.
\end{proof}

For any job $s$, define an \emph{$s$-schedule} to be a (partial)
schedule that schedules all jobs $j$ with $r_j\ge r_s$.  We use
notation $E_{s}$ to represent the minimum cost (energy) of an
$s$-schedule, including the cost of the possible gap between $r_{s}$
and its first block.


\begin{lemma}[partitioning]\label{lem: C partitioning}
There exists an optimal $s$-schedule $S$ with the following property:
Either $S$ does not have any small gap, or if $[u,t)$ is the first small
gap in $S$ and $h$ the number of gaps in $[r_s,u)$, then $u = U_{s,n,h}$.
\end{lemma}

\begin{proof}
Let $S$ be an optimal schedule. If $S$ does not have any small gaps,
we are done. Otherwise, let $[u,t)$ be the first small gap in $S$ and
let $\calJ$ be the set of jobs released in $[r_{s},u)$.
By Lemma~\ref{dominated_schedules}, we can assume that
all jobs from $\calJ$ are completed in $S$ no later than at time $u$.
This means that the segment of $S$ in $[r_s,u)$ is an
$(s,n)$-schedule, and thus $u\leq U_{s,n,h}$.

Towards a proof by contradiction, assume that this inequality is
strict, that is $u < U_{s,n,h}$.  We now use Lemma~\ref{lem: compression}
(the compression lemma).
First we show that the assumptions of this lemma are satisfied. By
Lemma~\ref{dominated_schedules}, no job is released in $[u,t)$, and
every job $j$ released before $u$ is completed in $S$ not later than at $u$, 
so $\greedyC_{j,n}\leq u$. Let $Q$ be the $(s,n)$-schedule with at most $h$ gaps
and completion time $U_{s,n,h}$. Now, applying Lemma~\ref{lem: compression}
    with $k'=n$ and $\theta = u$, we obtain that there is an $(s,n)$-schedule $R$,
 scheduling
    all jobs from $\calJ$, with completion time $v=\makespan(R)\leq t$
    and at most $h$ gaps. Moreover, if $v\leq u$ then $R$ has in fact at
    most $h-1$ gaps.

 We replace the segment of $S$ in $[r_s,t)$ by $R$, obtaining an
   $s$-schedule $S'$.  To complete the proof, it is sufficient to show
   that the cost of $S'$ is strictly smaller than that of $S$, as this
will contradict the optimality of $S$.  Schedules
$S$ and $S'$ are identical in $[t,\infty)$.  The cost of 
   the gaps of $S$ in $[r_s,t)$ is $Lh+t-u$.  If $v>u$, then the
     gaps in $S'$ in $[r_s,t)$ cost at most $Lh+t-v$, and if
       $v\leq u$, they cost at most $L(h-1)+L$, since the gap between
       $v$ and $t$ can cost at most $L$. Thus in both cases the cost of these
gaps is strictly smaller than $Lh+t-u$.
\end{proof}


\paragraph{Algorithm~{\algC}.}
The algorithm first computes the table $U_{s,k,g}$, for all  $s = 1,...,n$, $k =
0,...,n$, and $g = 0,1,...,n-1$, using either
Algorithm~{\algA} or {\algB}, whichever applies.  Then we use dynamic
programming to compute all values $E_{s}$.
These values will be stored in table $\EE{s}$ and computed
in order of decreasing release times $r_{s}$:
\begin{eqnarray} \label{decomp_C}
\EE{s} &\assign&
  \min_{0\le g\le n-1}\left\{\begin{array}{ll}
      Lg
              & \text{if } U_{s,n,g}>\max_{j} r_{j}
\\
      Lg+r_{l}-u+\EE{l}
              & \text{otherwise, where } u=U_{s,n,g},  r_l =\min\braced{ r_j \suchthat r_j > u}
             \end{array}\right.
\end{eqnarray}
The algorithm outputs $\EE{1}$ as
the minimum energy of the whole instance, where $r_{1}$ is
the first release time.  (Recall that the job $1$ is assumed to be
tight, so the schedule realizing $E_1$ will not have a gap at the
beginning.)

Note that the minimum (\ref{decomp_C}) is well-defined, for if
$u = U_{s,n,g} \le \max_{j} r_{j}$, then the frugality of the
schedule realizing $U_{s,n,g}$ implies that we have, in fact,
$u < \max_{j} r_{j}$, and therefore there is $l$ with $r_l > u$.

We now prove the correctness of Algorithm~{\algC} and analyze its
running time.


\begin{lemma}[feasibility of {\algC}]\label{lem: feasibility C}
For each job $s = 1,2,...,n$, we have $\EE{s}\ge E_s$.
\end{lemma}

\begin{proof}
We need to show that for each $s$ there is 
an $s$-schedule $S_s$ of cost at most $\EE{s}$.
The proof is by backward induction on $r_s$. 
In the base case, when $s$ is the job with
maximum release time, then we take $S_s$ to be the schedule
that executes $s$ at $r_s$. The cost of $S_s$ is $0$, so the
lemma holds.

Assume now that for any $s' > s$ we have already constructed an
$s'$-schedule $S_{s'}$ of cost at most $\EE{s'}$. Let $g$ be the value
that realizes the minimum in (\ref{decomp_C}). We distinguish two
cases, depending on which option realizes the minimum.

Suppose first that $\EE{s} = Lg$ and $U_{s,n,g}>\max_{j} r_{j}$.
Then there is a schedule of all jobs released at or after $r_s$ with at
most $g$ gaps. Let $S_s$ be this schedule. Since each gap's cost is at
most $L$, the total cost of $S_s$ is at most $Lg$.

The second case is when $\EE{s} = Lg + r_l - u + \EE{l}$,
where $u = U_{s,n,g} \le \max_{j} r_{j}$ and 
$r_l =\min\braced{ r_j \suchthat r_j > u}$.
Choose an $(s,n)$-schedule $Q$ with at most $g$ gaps and completion time $u$.
As explained right after the algorithm, 
the frugality of $Q$ implies that there is no job released at $u$, and
thus $l$ is well-defined.

By induction, there exists
an $l$-schedule $S_l$ of cost at most $\EE{l}$. We then define $S_s$ as the
disjoint union of $Q$ and $S_l$. The cost of $Q$ is at most $Lg$. 
If $v\ge r_l$ is the first start time of a job in $S_l$, write $\EE{l}$ as 
$\EE{l} = \min\braced{v-r_l,L} + E'$.  In other words, $E'$ is the cost of
the gaps in $S_l$ excluding the gap $[r_l,v)$ (if $r_l < v$).
Then the cost of $S_s$ is at most 
$Lg + \min\braced{v-u,L} + E' 
	\le Lg + (r_l - u) + \min\braced{v-r_l,L} + E' 
	= Lg + r_l - u + \EE{l} = \EE{s}$.
\end{proof}


\begin{lemma}[optimality of {\algC}]\label{lem: optimality C}
For each job $s = 1,2,...,n$, we have $\EE{s}\le E_s$.
\end{lemma}

\begin{proof}
For any job $s$, we now prove that any
$s$-schedule $S$ has cost at least $\EE{s}$. The proof is by backward
induction on $r_s$. In the base case, when
$s$ is the job that is released last, then $U_{s,n,0}>r_{s} = \max_j r_j$, 
so we have $\EE{s}=0$, and the lemma holds.

Suppose now that $s$ is a job that is not released last and let $S$ be an optimal
$s$-schedule. Without loss of generality, we can assume that $S$ satisfies 
Lemma~\ref{dominated_schedules} and Lemma~\ref{lem: C partitioning}.

If $S$ does not have any small gaps then, denoting by $g$ the number
of gaps in $S$, the cost of $S$ is exactly $Lg$. The existence of $S$
implies that $U_{s,n,g} > \max_jr_j$, so $\EE{s} \le Lg$, completing
the argument for this case.

Otherwise, let $[u,t)$ be the first small gap in $S$.
Denote by $S'$ the segment of $S$ in $[r_{s},u)$ and by $S''$ the segment of 
$S$ in $[t,\makespan(S))$. 
By Lemma~\ref{dominated_schedules}, $S''$ contains
only jobs $j$ with $r_{j}\ge t$.  In particular the job $l$ to be
scheduled at $t$ is released at $r_{l}=t$. Therefore $S''$ is an
$l$-schedule, and, by induction, we obtain
that the cost of $S''$ is at least $\EE{l}$.

Let $g$ be
number of gaps in $S'$. By Lemma~\ref{lem: C partitioning} we have 
$u=U_{s,n,g}$. So the cost of $S$ is $Lg + r_l - u + \EE{l} \ge \EE{s}$,
where the inequality holds because $u$, $g$ and $l$ satisfy the
condition in the second option of (\ref{decomp_C}). This completes
the proof.
\end{proof}


\begin{theorem}\label{thm: algorithm C}
Algorithm~{\algC} correctly computes the optimum solution for
$1|r_j|E$, and it can be implemented in time $O(n^5)$. Further, in the
special case $1|r_j; p_j = 1|E$, it can be implemented in time
$O(n^4)$.
\end{theorem}

\begin{proof}
The correctness of {\algC} follows from
Lemma~\ref{lem: feasibility C} and Lemma~\ref{lem: optimality C},
so it is sufficient to justify the time bound.
By Theorem~\ref{thm: algorithm A} and Theorem~\ref{thm: algorithm B}, 
we can compute the table
$U_{s,k,g}$ in time $O(n^4)$ and $O(n^5)$ for unit jobs and arbitrary
jobs, respectively. The post-processing, that is computing all values
$E_s$, can be easily done in time $O(n^2\log n)$, since we have $n$
values $E_s$ to compute, for each $s$ we minimize over $n-1$ values of
$g$, and for fixed $s$ and $g$ we can find the index $l$ in time
$O(\log n)$ with binary search.  (Finding this $l$ can be in fact reduced 
to amortized time $O(1)$ if we process $g$ in increasing order, for then
the values of $U_{s,n,g}$, and thus also of $l$, increase
monotonically as well.)
\end{proof}


\section{Final Comments}
\label{sec: Final Comments}


We presented an $O(n^5)$-time algorithm for the minimum energy
scheduling problem $1|r_j;{\pmtn}|E$, and an $O(n^{4})$ algorithm for
$1|r_j;p_{j}=1|E$.

Many open problems remain. Can the running times be improved further?
In fact, fast --- say, $O(n\log n)$-time --- algorithms with low
approximation ratios may be of interest as well.

For the multiprocessor
case, we are given $m$ parallel machines, and every job $j$ has to be assigned 
to $p_{j}$ time slots in $[r_{j},d_{j})$ which may belong to different machines. 
At any time a job can be scheduled on at most one machine. 
The goal is to minimize the total energy usage over all machines.
In \cite{DemaineGhodsi:Scheduling-to-minimize} an $O(n^{7}m^{5})$-time algorithm 
was given for this problem, for the special case when $L=1$ and the jobs have unit length. 
It would be interesting to extend the results of this paper to the multiprocessor case,
improving the running time and solving the general case for arbitrary $L$.

Another generalization is to allow multiple power-down states
\cite{IraniPruhs:Algorithmic-problems,IrShGu03A,LiYao:An-Efficient-Algorithm}.  Can this problem be
solved in polynomial-time?  In fact, the SS-PD problem discussed by
Irani and Pruhs in their survey \cite{IraniPruhs:Algorithmic-problems}
is even more general as it involves speed scaling in addition to
multiple power states, and its status remains open as well.



\bibliographystyle{plain}
\bibliography{energy}

\end{document}